\documentclass[pdflatex,sn-mathphys-num]{sn-jnl}

\usepackage{graphicx}%
\usepackage{multirow}%
\usepackage{amsmath,amssymb,amsfonts}%
\usepackage{amsthm}%
\usepackage{mathrsfs}%
\usepackage[title]{appendix}%
\usepackage{xcolor}%
\usepackage{textcomp}%
\usepackage{manyfoot}%
\usepackage{booktabs}%
\usepackage{algorithm}%
\usepackage{algorithmicx}%
\usepackage{algpseudocode}%
\usepackage{listings}%

\newcommand{\dd}{\mathinner{.\,.}}

\newcommand{\no}[1]{}

\newcommand{\Diego}[1]{\textcolor{red}{Diego: #1}}

\newcommand{\pending}[1]{\textcolor{red}{Pending: #1}}

\usepackage{booktabs}
\usepackage{pgfplots}
\usepackage{rotating}
\usepackage{multirow}
\usepackage{pdflscape}
\usepackage{soul}
\usepackage{titlesec}
\usepackage{caption}
\usepackage{subcaption}
\usepackage{array}
\usepackage{algorithm}
\usepackage{tabularx}
\usepackage{comment}
\usepackage{etoolbox}
\usepackage{xspace}
\usepackage{amsmath}
\usepackage{balance}

\newcommand{\dom}{\mathrm{dom}}
\newcommand{\var}{\text{vars}}
\newcommand{\U}{\mathcal{U}}
\newcommand{\V}{\mathcal{V}}

\newcommand{\leap}[1]{\leapname\mathsf{(}#1\mathsf{)}}

\newcommand{\leapname}{\mathsf{leap}}

\renewcommand{\log}{\lg}

\renewcommand{\to}{\textsc{o}}
\newcommand{\ts}{\textsc{s}}
\newcommand{\tp}{\textsc{p}}

\newcommand{\agm}[1]{\ensuremath{#1^*}}
\newcommand{\rank}{\mathsf{rank}}
\newcommand{\select}{\mathsf{select}}
\newcommand{\selectnext}{\mathsf{selectnext}}
\newcommand{\access}{\mathsf{access}}
\newcommand{\degree}{\mathsf{degree}}
\newcommand{\child}{\mathsf{child}}
\newcommand{\bit}[1]{\mathsf{#1}}
\newcommand{\rangenext}{\mathsf{range\_next\_value}}
\newcommand{\rangecount}{\mathsf{range\_count}\xspace}
\newcommand{\rangelist}{\mathsf{range\_count}\xspace}    
\newcommand{\rangeint}{\mathsf{range\_intersect}\xspace}
\newcommand{\LTJ}{\textrm{LTJ}\xspace}

\newcommand{\rdfcsa}{{\it rdfcsa}\xspace}
\newcommand{\rdfcsaspo}{{\it rdfcsa$^{\textsc{\sc spo}}$}\xspace}
\newcommand{\rdfcsaops}{{\it rdfcsa$^{\textsc{\sc ops}}$}\xspace}
\newcommand{\csa}{{\rm CSA}\xspace}

\newcommand{\rangeCsa}{\mathsf{range}}
\newcommand{\downCsa}{\mathsf{down}}

\newcommand{\findTargetPsiCsa}{\mathsf{findTarget_{\Psi}}}
\newcommand{\findTargetPsiPsiCsa}{\mathsf{findTarget_{\Psi\Psi}}}
\newcommand{\limitV}{\mathsf{limitV}}
\newcommand{\ring}{{\it ring}\xspace}

\newcommand{\qdag}{{\it qdag}\xspace}
\newcommand{\qdags}{{\it qdags}\xspace}
\newcommand{\compactLTJ}{{\it compactLTJ}\xspace}
\newcommand{\compactLTJstar}{{\it compactLTJ*}\xspace}
\newcommand{\unCompactLTJ}{{\it unCompactLTJ}\xspace}
\newcommand{\CompactLTJ}{$\mathtt{CLTJ}$\xspace}
\newcommand{\CompactLTJstar}{$\mathtt{CLTJ*}$\xspace}
\newcommand{\UnCompactLTJ}{$\mathtt{UnCLTJ}$\xspace}
\newcommand{\UnCompactLTJstar}{$\mathtt{UnCLTJ*}$\xspace}
\newcommand{\Ring}{$\mathtt{Ring}$\xspace}
\newcommand{\Rdfcsa}{$\mathtt{RDFCSA}$\xspace}
\newcommand{\URing}{$\mathtt{URing}$\xspace}
\newcommand{\VRing}{$\mathtt{VRing}$\xspace}
\newcommand{\VURing}{$\mathtt{VURing}$\xspace}
\newcommand{\IRing}{$\mathtt{IRing}$\xspace}
\newcommand{\IURing}{$\mathtt{IURing}$\xspace}

\newcommand{\QDags}{\texttt{QDags}\xspace}
\newcommand{\QDag}{\texttt{QDag}\xspace}

\newcommand{\blockcomment}[1]{}

\theoremstyle{thmstyleone}%
\theoremstyle{thmstyletwo}%

\theoremstyle{thmstylethree}%

\begin{document}

\title[New Compressed Indices for Multijoins on Graph Databases]{New Compressed Indices for Multijoins on Graph Databases}


\author[1,4]{\fnm{Diego} \sur{Arroyuelo}}\email{diego.arroyuelo@uc.cl}
\equalcont{These authors contributed equally to this work.}

\author[2,4]{\fnm{Fabrizio} \sur{Barisione}}\email{fbarisio@dcc.uchile.cl}
\equalcont{These authors contributed equally to this work.}

\author[3,5]{\fnm{Antonio} \sur{Fariña}}\email{antonio.farina@udc.es}
\equalcont{These authors contributed equally to this work.}

\author*[3,4]{\fnm{Adrián} \sur{Gómez-Brandón}}\email{adrian.gbrandon@udc.es}
\equalcont{These authors contributed equally to this work.}

\author[2,4]{\fnm{Gonzalo} \sur{Navarro}}\email{gnavarro@dcc.uchile.cl}
\equalcont{These authors contributed equally to this work.}

\affil[1]{\orgname{Pontificia Universidad Católica}, \orgaddress{\street{Av. Vicuña Mackenna}, \city{Santiago}, \country{Chile}}}

\affil[2]{\orgname{Universidad de Chile}, \orgaddress{\street{Beauchef}, \city{Santiago}, \country{Chile}}}

\affil[3]{\orgname{Universidade da Coruña}, \orgaddress{\street{Campus de Elviña}, \city{A Coruña}, \country{Spain}}}

\affil[4]{\orgname{IMFD}, \orgaddress{\street{Av. Vicuña Mackenna}, \city{Santiago}, \country{Chile}}}

\affil[5]{\orgname{CITIC}, \orgaddress{\street{Campus de Elviña}, \city{A Coruña}, \country{Spain}}}



\abstract{A recent surprising result in the implementation of worst-case-optimal (wco) multijoins in graph databases (specifically, basic graph patterns) is that they can be supported on graph representations that take even less space than a plain representation, and orders of magnitude less space than classical indices, while offering comparable performance. In this paper we uncover a wide set of new wco space-time tradeoffs: we (1) introduce new compact indices that handle multijoins in wco time, and (2) combine them with new query resolution strategies that offer better times in practice. As a result, we improve the average query times of current compact representations by a factor of up to 13 to produce the first 1000 results, and using twice their space, reduce their total average query time by a factor of 2. Our experiments suggest that there is more room for improvement in terms of generating better query plans for multijoins.}




\maketitle

\section{Introduction}

Natural joins are fundamental in the relational algebra, and generally the most costly operations. A bad implementation choice can lead to unaffordable query times, so they have been a concern since the beginnings of the relational model. Apart from efficient algorithms to join two tables (i.e., solve pair-wise joins), database management systems sought optimized strategies (e.g., \cite{Sel79}) to solve joins between several tables (i.e., multijoins),  as differences between good and bad plans could be huge in terms of efficiency. A {\em query plan} for a multijoin was a binary expression tree where the leaves were the tables to join and the internal nodes were the pair-wise joins to perform. 

After half a century of revolving around this pairwise-join-based strategy, it was found that it had no chance to be optimal \cite{AGM13}, as it could generate intermediate results (at internal nodes of the expression tree) that were much larger than the final output. The concept of a {\em worst-case optimal (wco)} algorithm \cite{AGM13} was coined to define a multijoin algorithm taking time $\tilde{O}(Q^*)$, where $Q^*$ is the largest output size on some database instance with the same table sizes of the given one ($\tilde{O}(Q^*)$ allows multiplying $Q^*$ by terms that do not depend, or depend only logarithmically, on the database size). Several wco join algorithms were proposed since then \cite{NPRR12,NRR13,leapfrog,geometric,NABKNRR15,tutorialngo}. 

{\em Leapfrog Triejoin (LTJ)} \cite{leapfrog} is probably the simplest and most popular wco algorithm. At a high level, it can be regarded as reducing the multijoin by one {\em attribute} at a time, instead of by one {\em relation} at a time as in the classical query plans. LTJ chooses a suitable order in which the joined attributes will be {\em eliminated} (which means finding all their possible values in the output and branching on the subset of the output matching each such value). To proceed efficiently, LTJ needs the rows of each relation stored in a trie (or digital tree) where the root-to-leaf attribute order is consistent with the chosen attribute elimination order. Even though LTJ is wco with any elimination order, it turns out that, just like with the traditional query plans, there can be large performance differences when choosing different orders \cite{leapfrog,HRRSiswc19}. This means, first, that choosing a good order is essential and, second, that LTJ needs tries storing each relation {\em in every possible order of its attributes}, that is, $d!$ tries for a relation with $d$ attributes.

This high space requirement shows up, in one form or another, in all the existing wco algorithms, and has become an obstacle to their full adoption in database systems. Wco algorithms are of particular interest in {\em graph databases}, which can be regarded as labeled graphs, or as a single relational table with three attributes: source node, label, and target node. Standard query languages for graph databases like SPARQL \cite{sparql11} feature most prominently {\em basic graph patterns (BGPs)}, which essentially are a combination of multijoins and simple selections. The concept of wco algorithms, as well as LTJ, can be translated into solving BGPs on graph databases \cite{HRRSiswc19}. This is very relevant because typical BGPs correspond to large and complex multijoins \cite{NABKNRR15,EmptyHeaded,KalinskyEK17,HRRSiswc19}, where non-wco algorithms can be orders of magnitude slower than wco ones \cite{EmptyHeaded}. Still, LTJ needs $3! = 6$ copies of the database in the form of tries, which even for this low arity is sufficiently space-demanding to discourage its full implementation.

The implementation of various wco indices seems to confirm that large space usage will be the price for featuring wco query times. For example, a wco version of Jena \cite{HRRSiswc19} doubles the space of the original non-wco version. Efficient wco implementations like EmptyHeaded \cite{EmptyHeaded} and MillenniumDB \cite{VR+23} use many times the space required to store the raw data. Surprisingly, recent research debunks this impression. 
In particular, the {\em ring} \cite{AHNRRSsigmod21,AGHNRRStods24} is a novel compact index that represents graph databases (the data {\em and} the index structures) within {\em less} space than that used by the raw data in plain form, while still supporting BGPs within competitive times, often even lower than indices that are orders of magnitude larger. 

\subsection{Our contribution}
The unexpected result achieved by small indexes like the {\em ring} has opened numerous opportunities for new space-time tradeoffs in index data structures for wco multijoins on graph databases. The {\em ring} was aimed at minimum space usage, to demonstrate that competitive query times could be achieved using only as much space as the raw data, and even less. Since this space is much lower than that of traditional indices, there is sufficient slack to introduce larger data structures that, still using a fraction of the space of those traditional indices, are much faster than the \ring. Additionally, despite occupying minimal space, the data structures supporting the \ring enable efficient computation of information---which would otherwise need to be explicitly stored by conventional indices---that helps compute efficient attribute elimination orders for LTJ \cite{AGHNRRStods24}.
Motivated by this, we contribute with new compact indices that support solving BGPs in wco time, and their combination with new query resolution techniques, thereby uncovering a wide set of new space-time tradeoffs in wco indices for solving BGPs. Concretely:
\begin{enumerate}
\item We design an alternative to the \ring that, using twice its space, is four times faster in the median and twice as fast on the average. This new index, which we call the \rdfcsa, builds on an existing compact index representation that only supported single joins \cite{BCdBFNsupe22}, so that now it supports full BGPs in wco time.
\item We combine the \ring and the \rdfcsa with an {\em adaptive} variable elimination order, which recomputes the best elimination order as the join proceeds and more information is available. We use new estimators for the next variable to bind that are more accurate and can be computed efficiently on our compact indices. In our experiments, the combination obtains the first thousand results 4--13 times faster, on average, than the traditional global-order strategy. We show that our adaptive strategies outperform, in many cases, the {\em best possible} global-order strategy.
\end{enumerate}


%
%
%
%

\section{Preliminary concepts}

\subsection{Graph joins}

\subsubsection{Edge-Labeled Graphs}

Let $\U$ be a totally ordered, countably infinite set of {\em constants}, which we call the \emph{universe}. In the RDF model ~\cite{rdf}, an \emph{edge-labeled graph} is a finite set of {\em triples} $G \subseteq \U^3$, where each triple
$(s, p, o)\in\U^3$ encodes the directed edge $s \xrightarrow{p} o$ from vertex $s$ to vertex $o$, with edge label $p$. We call $\dom(G) = \{s,p,o~|~(s,p,o) \in G\}$ the subset of $\U$ used as constants in $G$. 
For any element $u\in\U$, let $u+1$ denote the successor of $u$ in the total order $\U$. We also denote $U = \max \dom(G)$. For simplicity, we will assume that the constants in $\U$ have been mapped to integers in the range $[1\dd U]$, and will even assume $\U = [1\dd U]$.


\subsubsection{Basic Graph Patterns (BGPs)}

A graph $G$ is often queried to find patterns of interest, that is, subgraphs of $G$ that are homomorphic to a given pattern $Q$. 
Unlike the graph $G$, which is formed only by constants in $\U$, a pattern $Q$ can contain also {\em variables}, formally defined as follows. 
Let $\V$ denote an infinite set of variables, such that $\U\cap\V=\emptyset$. Then, a \emph{triple pattern} $t$ is a tuple $(s,p,o) \in (\U\cup\V)^3$, and a \emph{basic graph pattern} is a finite set $Q \subseteq (\U\cup\V)^3$ of triple patterns.
Each triple pattern in $Q$ is an atomic query over the graph, equivalent to equality-based selections on a single ternary relation. Thus, a basic graph pattern (BGP) corresponds to a full conjunctive query (i.e., a \textit{join query} plus simple selections) over the relational representation of the graph. 

Let $\var(Q)$ denote the set of variables used in pattern $Q$. The \emph{evaluation} of $Q$ over a graph $G$ is then defined to be the set of mappings $Q(G) := \{ \mu : \var(Q) \rightarrow \dom(G) \mid \mu(Q) \subseteq G\}$, called \emph{solutions}, where $\mu(Q)$ denotes the image of $Q$ under $\mu$, that is, the result of replacing each variable $x \in \var(Q)$ in $Q$ by $\mu(x)$.


\subsection{Worst-case optimal joins}
\subsubsection{The AGM bound}
A well-established bound to analyze join algorithms is the \emph{AGM bound}, introduced by Atserias et al.~\cite{AGM13}, which  sets a limit on the maximum  output size for a natural join query. Let $Q$ denote such a query and $D$ a relational database instance. The AGM bound of $Q$ over $D$, denoted $\agm{Q}$, is the maximum number of tuples generated by evaluating $Q$ over any database instance $D'$ containing a table $R'$ for each table $R$ of $D$, with the same attributes and $|R'| \le |R|$ tuples. 
Though BGPs extend natural joins with self joins, constants in $\U$, and the multiple use of a variable in a triple pattern, the AGM bound can still be applied to them by regarding each triple pattern as a relation formed by the triples that match its constants \cite{HRRSiswc19}. 

Given a join query (or BGP) $Q$ and a database instance $D$, a {\em join algorithm} enumerates $Q(D)$, the solutions for $Q$ over $D$. A join algorithm is {\em worst-case optimal (wco)} if it has a running time in $\tilde{O}(Q^*)$, which is $O(Q^*)$ multiplied by terms that do not depend, or depend only polylogarithmically, on $|D|$. 
Atserias et al. \cite{AGM13}~proved that there are queries $Q$ for which no plan involving only pair-wise joins can be wco. 

This paper focuses on wco algorithms, precisely on the one described next, which is the one most frequently implemented.

\subsubsection{Leapfrog TrieJoin (LTJ)} 
\label{subsec:ltj}

We describe the Leapfrog Triejoin algorithm \cite{leapfrog}, originally designed for natural joins in relational databases, as it is adapted for BGP matching on labeled graphs \cite{HRRSiswc19}.

Let $Q = \{t_1, \ldots, t_q\}$ be a BGP and $\var(Q) = \{x_1, \ldots, x_v\}$ its set of variables. \LTJ uses a {\em variable elimination} approach, which extends the concept of attribute elimination. The algorithm carries out $v = |\var(Q)|$ iterations, handling one particular variable of $\var(Q)$ at a time. This involves defining a total order $\langle x_{i_1}, \ldots, x_{i_v} \rangle$ of $\var(Q)$, which we call a {\em VEO} for {\em variable elimination order}. 

\begin{figure}[t]
\begin{center}
    \includegraphics[width=0.7\textwidth]{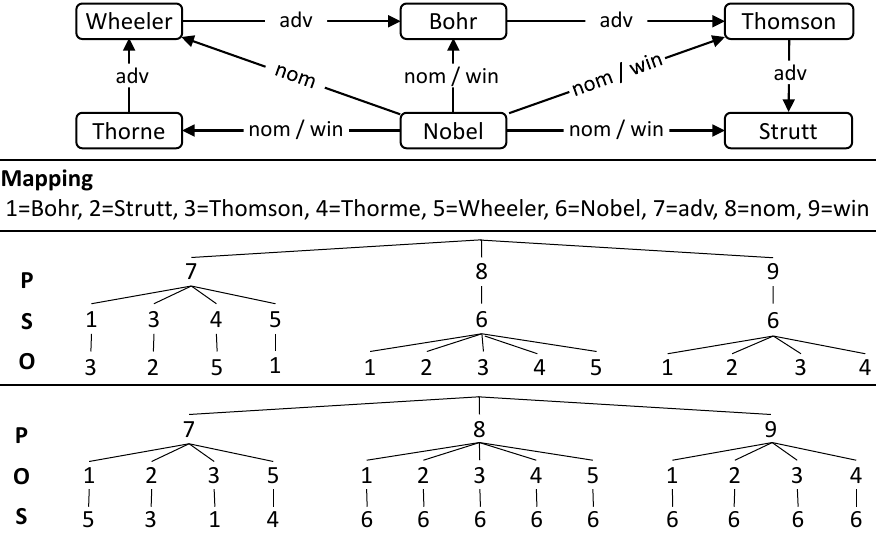}
    \caption{A labeled graph $G'$ with its string to integer mapping and tries for orders \textsc{pso} and \textsc{pos}.}
    \label{fig:graph_mapping_orders}
\end{center}
\end{figure}

Each triple pattern $t_i$ is interpreted as a relation that will be joined, and associated with a suitable trie $\tau_i$. The root-to-leaf path in $\tau_i$ must start with the constants that appear in $t_i$, and the rest of its levels must visit the variables of $t_i$ in an order that is consistent with the VEO chosen for $Q$ (this is why we need the $3!=6$ tries).
Fig.~\ref{fig:graph_mapping_orders} shows an example graph and the corresponding mapping of the constants in $\U$ to integers. We also show two tries representing the graph triples using the orders \textsc{pso} (i.e., predicate, subject, object) and \textsc{pos}. For example, we must use the trie \textsc{pso} to handle a triple pattern $(x,8,y)$ if the VEO is $\langle x,y \rangle$, and the trie \textsc{pos} if the VEO is $\langle y,x \rangle$. If $Q$ has a second triple pattern $(y,7,x)$, then we need both tries no matter the VEO we use.

 The algorithm starts at the root of every $\tau_i$ and descends by the children that correspond to the constants in $t_i$.
 We then proceed to the variable elimination phase. Let $Q_j \subseteq Q$ be the triple patterns that contain variable $x_{i_j}$. Starting with the first variable, $x_{i_1}$, \LTJ finds each $c\in\dom(G)$ such that for every $t \in Q_1$, if $x_{i_1}$ is replaced by $c$ in $t$, the evaluation of the modified triple pattern $t$ over $G$ is non-empty (i.e., there may be answers to $Q$ where $x_{i_1}$ is equal to $c$). If the trie $\tau$ of $t$ is consistent with the VEO, then the children of its current node contain precisely those suitable values $c$ for variable $x_{i_1}$.
 
During the execution, we keep a mapping $\mu$ with the solutions of $Q$. As we find each constant $c$ suitable for $x_{i_1}$, we {\em bind} $x_1$ to $c$, that is, we set $\mu = \{(x_1 := c)\}$ and branch on this value $c$. In this branch, we go down by $c$ in all the virtual tries $\tau$ such that $t \in Q_1$. We now repeat the same process with $Q_2$, finding suitable constants $d$ for $x_{i_2}$ and increasing the mapping to $\mu = \{(x_1 := c), (x_2 := d)\}$, and so on. Once we have bound all variables in this way, $\mu$ is a solution for $Q$ (this happens many times because we branch on every binding to $c$, $d$, etc.). When it has considered all the bindings $c$ for some variable $x_{i_j}$, \LTJ backtracks and continues with the next binding for $Q_{j-1}$. When this process finishes, the algorithm has reported all the solutions for $Q$. 

Operationally, the values $c$, $d$, etc.\ are found by {\em intersecting} the children of the current nodes in all the tries $\tau_i$ for $t_i \in Q_j$. \LTJ carries out the intersection using the primitive $\leap{\tau_i,c}$, which finds the next smallest constant $c_i\geq c$ within the children of the current node in trie $\tau_{i}$; if there is no such value $c_i$, $\leap{\tau_i,c}$ returns a special value $\perp$. 

\subsection{Variable Elimination Orders (VEOs)} \label{sec:veo}

Veldhuizen \cite{leapfrog} showed that if $\leap{}$ runs in polylogarithmic time, then \LTJ is wco no matter the VEO chosen, as long as the tries used have the right attribute order. In practice, however, the VEO plays a fundamental role in the efficiency of the algorithm \cite{leapfrog,HRRSiswc19}. 
 A VEO yielding a large number of intermediate solutions that are later discarded during \LTJ execution, will be worse than one that avoids exploring many such alternatives. One would prefer, in general, to first eliminate selective variables (i.e., the ones that yield a smaller candidate set when intersecting).

A heuristic to generate a good VEO in practice \cite{HRRSiswc19,AHNRRSsigmod21,VR+23} computes, for each variable $x_j$, its minimum weight 
\begin{equation} \label{eq:wj}
w_j = \min \{ w_{ij} ~|~ x_j \textrm{ appears in triple } t_i \},
\end{equation}
where $w_{ij}$ is the {\em weight} of $x_j$ in $t_i$. The VEO sorts the variables in increasing order of $w_j$, with a couple of restrictions: (i) each new variable should share some triple pattern with a previous variable, if possible; (ii) variables appearing only once in $Q$ (called {\em lonely}) must be processed at the end.

To compute $w_{ij}$, we (temporarily) choose a trie $\tau_j$ where $x_j$ appears right after the constants of $t_i$, and descend in $\tau_j$ by the constants. The number of children of the trie node $v$ we have reached is the desired weight $w_{ij}$. This is the size of the list in $\tau_i$ to intersect when eliminating $x_j$.

In this paper we explore the use of {\em adaptive} VEOs, which are defined progressively as the query processing advances, and may differ for each different binding of the preceding variables. ADOPT \cite{adopt} is the first system combining LTJ with adaptive VEOs. The next variables to bind are chosen using reinforcement learning, by partially exploring possibly upcoming orders, and balancing the cost of exploring with that of the obtained improvements. Our adaptive VEOs will be computed, instead, simply as a variant of the formula presented above for global VEOs \cite{HRRSiswc19}. 

We will also explore more refined estimations of $w_j$ in Eq.~(\ref{eq:wj}), beyond the use of simply the minimum of the set sizes $w_{ij}$ to estimate the size of their intersection.

\no{
\subsubsection{Quadtree.} \label{sec:quad}
A quadtree is a geometric structure used to represent data points in grids of size $l \times l$. Assume $l$ is a power of two, then a quadtree is recursively defined as follows: if $l = 1$, then the grid has only one cell, which may contain a point or not. For $l > 1$, if the grid has no points, then the quadtree is a leaf. Otherwise, the quadtree is an internal node with four quadtree children, one per $l/2 \times l/2$ quadrant of the grid. The sequence of children chosen, from root to leaf, when reading the quadtree path that represents a point, yields the consecutive bits of its two coordinates, which are thereby represented implicitly.

A compact representation of quadtrees \cite{BLNis13.2} is analogous to the LOUDS 
representation: the quadtree is traversed levelwise and each internal node is encoded with four bits, indicating with a \bit{0} which of its children are empty grids (and, thus, leaves) and with a \bit{1} which are nonempty (and, thus, internal nodes). In the final level, however, those four bits tell which cells are empty or nonempty. Formulas very similar to those of LOUDS allow traversing the edges
in constant time. Overall, $p$ points are encoded using $4p\log l + o(p\log l)$ bits, twice a plain storage, and less if the points are clustered.

Quadtrees can be generalized to higher dimensions $d$. There exists a representation \cite{BDMRRR05} that requires only $(d + 2)p\log l + o(p \log l) + O(\log d)$ bits, also supporting edge traversal in $O(1)$ time. This poses a constant additive space overhead per point, independent of $d$. 
}

\no{
\section{Wco joins in compact space}
\label{sec:wco_cds}

In this section we describe the two main existing ways to implement wco join algorithms in compact space.

\subsection{QDags} \label{subsec:qdags}

\QDags \cite{ANRRtods22} represent each relation $R$ with attributes $A_1,\ldots,A_d$ as a $d$-dimensional grid with $|R|$ points, one per tuple in $R$. The value of each attribute $A_i$ is the $i$th coordinate of the corresponding point. The grids are represented as compressed quadtrees (Section~\ref{sec:quad}).

Since a join query is essentially an intersection on the common attributes of the involved relations, it is carried out using \QDags in two steps: $(i)$ the joined relations are {\em extended} to contain all the $d$ attributes of the output relation (i.e., the union of the attributes of the joined relations), and $(ii)$ the extended relations are traversed in synchronization to find the points that exist in all of them. 

The extension of $R$ implies copying every point in $R$ to new points in its extension, one for every possible combination of the new attributes. For example, to extend a relation $R(A,B)$ to $R'(A,B,C)$, for each point $(a,b) \in R$ we must create the points $(a,b,c)$ for every $c$. While doing this physically is impractical, it is simulated with just $O(d)$ additive extra space by creating a {\em mapping} vector that remaps the extended coordinates to the original ones. For example, let $R(A,B)$ be represented as a 2-dimensional quadtree, where nodes have 4 children. The grid of $R'$ has a line of points, along the coordinate for attribute $C$, for each point $(a,b)$ in the grid of $R$. To represent $R'$, we simulate a virtual 3-dimensional quadtree, where nodes have 8 children. A mapping $M=\langle 1,2,3,4,1,2,3,4 \rangle$ indicates that the contents of the 4 octants on the back are the same of the 4 octants on the front. That is, the first 4 octants of $R'$ are the 4 quadrants of $R$, and the last 4 octants of $R'$ are also the 4 quadrants of $R$. Repeating this interpretation in every subtree, we lift the dimension of a relation in just $O(d)$ time and extra space, while traversing edges in constant time. The structure formed by a quadtree and a dimension-lifting mapping is called a \qdag.

The second step is a simple traversal, in synchronization, of all the \qdags representing the extended relations, while creating the output quadtree $Q$. Every node where some \qdag has a leaf implies that $Q$ also has a leaf and we do not recurse further. Otherwise, we create an internal node in $Q$ and recursively enter into the first to the $2^d$th children of the joined \qdags. It might be that, after returning from the recursion, all those children are empty, in which case the current node of $Q$ must become a leaf too. This is why the algorithm cannot guarantee to take time proportional to the size of its output. It can proved to be, however, wco \cite{ANRRtods22}, while representing each relation $R$ using only {\em one} quadtree.

The price for supporting higher dimensions $d$ shows up in the time, however, where a term of the form $2^d$ multiplies the wco time complexity. This shows up in their experiments, where \QDags excel in multijoins producing tables of $d=3$ or $d=4$ attributes, but quickly worsen afterwards.

In a sense, the quadtree representation can be seen as a form of trie, where instead of considering (all the bits) of one attribute, then a second attribute, and so on, we consider the first bit of all attributes, then the second bit of all attributes, and so on. This allows supporting any join query in wco time with a single trie.

\blockcomment{
Fig.\ref{fig:quadtree_ex} shows an example with points in R, $P_1=(3,4)$ and $P_2=(8,6)$, coded as: "011100" and "111011", respectively. 
\begin{figure}[t]
\begin{center}
\includegraphics[scale=0.5]{img/quadtree.png}
\caption{Quadtree example, the black cells represents the points $P_1=(3,4)$ and $P_2=(8,6)$, coded as: "011100" and "111011", respectively.}
\label{fig:quadtree_ex}
\end{center}
\end{figure}
}

\blockcomment{
\subsubsection{Indexing the data.}

\pending{this.}

\subsubsection{Trie navigation.}

\pending{this.}
}

}

\section{The Ring: Wco joins in compact space} \label{sec:ring}

The \ring \cite{AHNRRSsigmod21,AGHNRRStods24} is an index that 
supports the 6 orders needed by \LTJ using a single data structure that uses space close to the raw data representation (and possibly less), while supporting the $\leap{}$ operation on the tries in logarithmic time.

\subsection{Bitvectors and wavelet trees} \label{sec:bv}
We start surveying the compact data structures used by the \ring. First, 
a {\em bitvector} $B[1\dd n]$ is an array of $n$ bits supporting the following queries:
\begin{itemize}
    \item $\access (B, i)$: the bit stored at $B[i]$.
    \item $\rank_b(B, i)$: the number of bits $b \in \{\bit{0},\bit{1}\}$ in $B[1\dd i]$.
    \item $\select_b(B, j)$ : the position of the $j$th occurrence of bit $b \in \{\bit{0},\bit{1}\}$ in $B$.
    \item $\selectnext_b(B,j)$ : the position of the leftmost occurrence of $b$ in $B[j\dd n]$.
\end{itemize}
These operations can be supported in $O(1)$ time using $n+o(n)$ bits of space \cite{Cla96,Mun96} or, alternatively, $nH_0(B)+o(n)$ bits \cite{RRR07}, where $H_0(B)\le 1$ denotes the zero-order entropy of $B$.

\label{subsec:wt}

The wavelet tree \cite{GGV03,Nav12} is a binary tree that represents a string $S[1\dd n]$ of symbols from an alphabet $\Sigma = \{1,\ldots, \sigma\}$. Each node $v$ represents a range $[a, b]$ of the alphabet, $a, b \in \Sigma$, with the root representing the whole alphabet $[1,\sigma]$ and each leaf representing a single symbol $a$, or range $[a,a]$. The range $[a,b]$ of internal nodes is divided into two, $[a,\lfloor (a+b)/2\rfloor]$ and $[\lfloor (a+b)/2 \rfloor+1,b]$, which are those of their left and right children.

Each internal node $v$ representing a range $[a,b]$ is associated with the subsequence $S_{a,b}$ of $S$ formed by the symbols in $[a,b]$ ($S_{a,b} = S$ if the node is the root). Instead of storing $S_{a,b}$, the node  stores a bitvector $B_{a,b}[1\dd|S_{a,b}|]$, where $B_{a,b}[i] = \bit{0}$ iff $S_{a,b}[i] \in [1,\lfloor (a+b)/2\rfloor]$ (i.e., belongs to the first half of the alphabet range); else $B_{a,b}[i] = \bit{1}$.

Note that the bitvector lengths at any level of the tree sum up to $n$ and we need to support binary $\rank$/$\select$ operations on them. Therefore, the wavelet tree represents $S$ using $n\log\sigma+o(n\log\sigma)$ bits (a plain representation uses almost the same, $n\log\sigma$ bits), and even within zero-order entropy, $nH_0(S) \le n\log\sigma$ bits. For large alphabets (as occurs in this paper), the additional space for the $O(\sigma)$ tree pointers are eliminated in a pointerless version called wavelet matrix \cite{CNO15}.

\begin{figure}[t]
\centering
\includegraphics[width=0.7\textwidth]{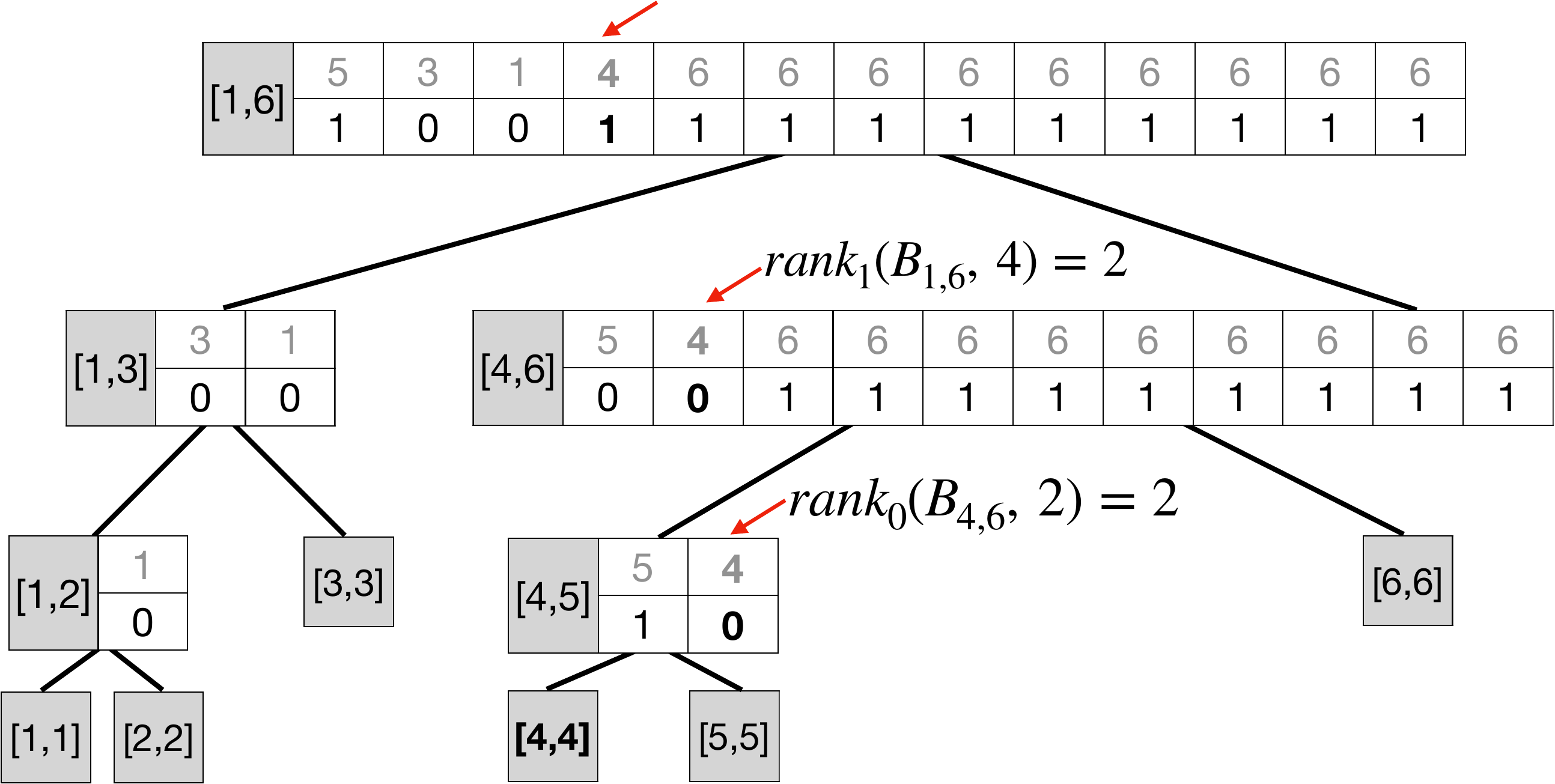}
\caption{Example of the wavelet tree for the sequence $\{5,3,1,4,6,6,6,6,6,6,6,6,6\}$. The ranges at the left depict the alphabet range of each bitmap. The arrows show the procedure to obtain the $4$th value of the sequence. 
}
\label{fig:wt}
\end{figure}

The wavelet trees support the functionality of $\access$, $\rank$, and $\select$ on general alphabets in time $O(\log\sigma)$ by traversing the tree from the root to a leaf. For instance, to access $S[i]$ we start at position $i$ in the bitmap of the root $B_{1,\sigma}$. Depending on $B_{1,\sigma}[i]$ we know that $S[i]$ is represented in the left (0) or right child (1). Hence, we continue by the left (resp. right) child at position $rank_0(B_{1,\sigma}, i)$ (resp., $rank_1(B_{1,\sigma}, i)$) when $B_{1,\sigma}[i]=0$ (resp., $B_{1,\sigma}[i]=1$). Those steps are repeated recursively within the corresponding bitmaps up to reaching a leaf. The symbol of that leaf is the solution to $S[i]$. Fig.~\ref{fig:wt} shows an example of $\access$ operation at position $4$. {Operation  $\rank$ is solved analogously, and $\select$ involves a further bottom-up traversal using $select$ on the bitmaps.

In addition, the wavelet trees support the following advanced operations that are useful for the \ring \cite{GNP11,BCN13}:
\begin{itemize}
\no{
\item $\access$, $\rank$, and $\select$, which extend their binary version to general alphabets, in time $O(\log\sigma)$.}
\item $\rangenext(S, r_s, r_e, c)$: 
for $c \in \Sigma$, finds in time $O(\log\sigma)$ the smallest symbol $c' \geq c$ that occurs within $S[r_s\dd r_e]$. This is used to simulate the primitive $\leap{}$ of \LTJ on a compact representation of $G$.

\item $\rangeint(S_1\langle [l_1, r_1],\ldots, S_k[l_k, r_k]\rangle)$: computes the intersection of the ranges $S_1[l_1\dd r_1], \ldots, S_k[l_k\dd r_k]$, reporting the symbols that occur in all the $k$ ranges. It is assumed that all the sequences $S_i$ share the same alphabet. This intersection is typically faster than the one performed via $\leap{}$. 
\item $\rangelist(S, x_s, x_e, [r_s, r_e]))$: counts how many symbols in $S[r_s\dd r_e]$ belong to the range $[x_s, x_e]$ in $O(\log \sigma)$ time. This will be used to estimate the costs of different VEOs on compressed representations of $G$.
\end{itemize}

\subsection{Indexing the data}
To represent a labeled graph $G$, let us define the table $T_{\textsc{spo}}[1\dd n][1\dd 3]$ storing the $n$ graph triples sorted according to the \textsc{spo} order. Column 1 of $T_{\textsc{spo}}$ corresponds to \ts, column 2 to \tp, and column 3 to \to. We denote $C_\to$ the last column of $T_{\textsc{spo}}$. 
Indeed, column $C_\to$ reads in left-to-right order the last level (i.e., the one corresponding to \to) of the trie for \textsc{spo}. Next, the process moves column $C_\to$ to the front in $T_{\textsc{spo}}$, making it the first column. The table is then sorted to obtain table $T_{\textsc{osp}}$, which conceptually represents the trie for the order \textsc{osp}. Let $C_\tp$ denote the last column of this table. Finally, column $C_\tp$ is moved to the front of $T_{\textsc{osp}}$ and the table is sorted again, obtaining table $T_{\textsc{pos}}$ and column $C_\ts$.
See Fig.~\ref{fig:ring}.

\begin{figure}[t]
\centering
\includegraphics[width=0.7\textwidth]{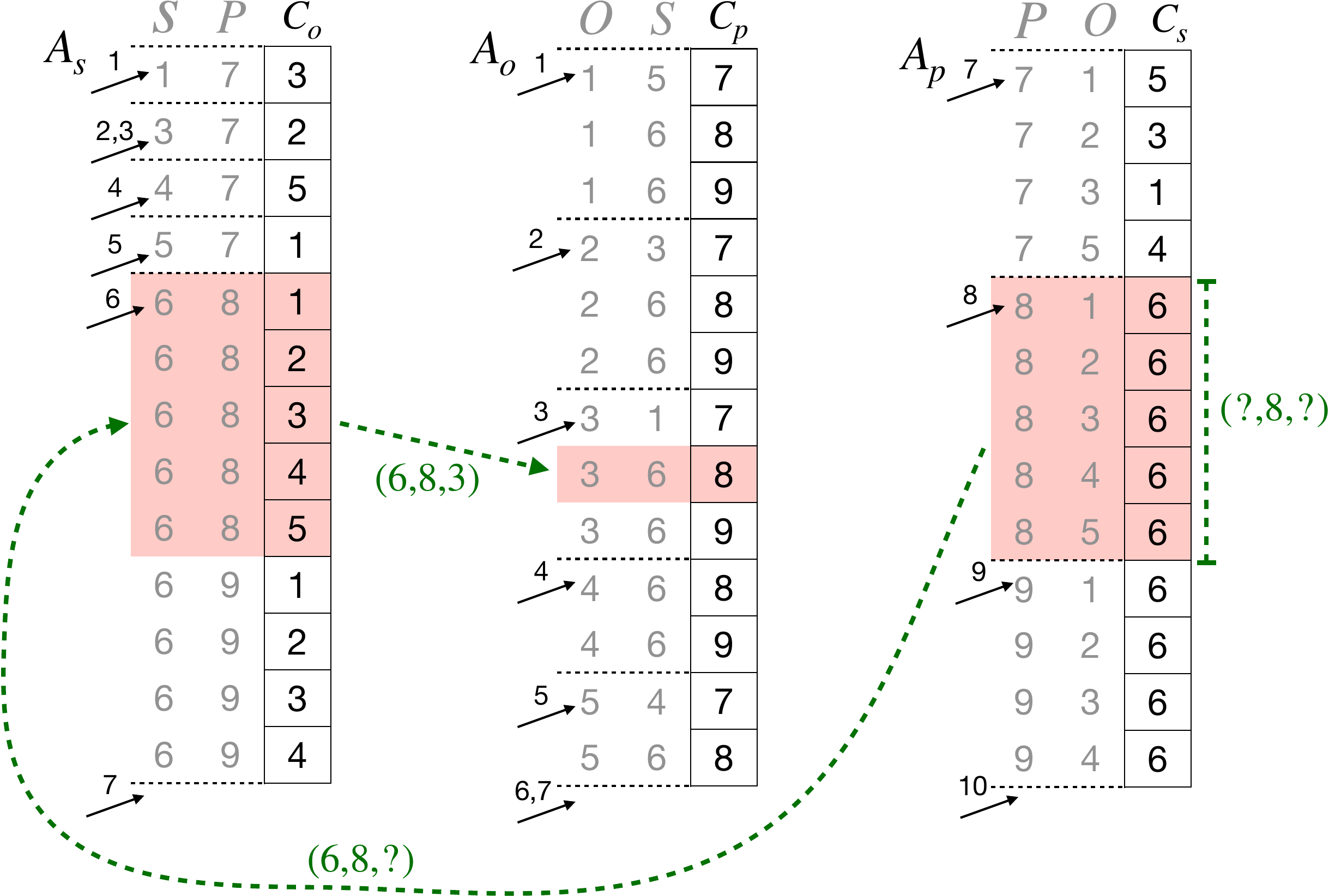}
\caption{The \ring representation of the graph of Fig.~\ref{fig:graph_mapping_orders}. The horizontal lines mark the values of $A_\ts$, $A_\to$, $A_\tp$, left to right.}
\label{fig:ring}
\end{figure}

The ring index is then formed by the sequences $C_*$, which are stored using wavelet trees (Section~\ref{subsec:wt}), with a total space requirement of $3n\lg{U} + o(n\lg{U})$ bits. We also build arrays $A_j$, for each $C_j$ with $j \in \{\ts, \tp, \to\}$, defined as $A_j[k] = |\{i\in[1\dd n],~C_j[i] < k\}|$, for $k = 1,\ldots, U+1$. These arrays store the cumulative number of occurrences of the symbols of $\U$ in $C_j$. This adds $O(U\log n)$ extra bits, which are $o(n\log U)$ if $U \in o(n)$. In practice, these arrays are represented using bitvectors (Section~\ref{sec:bv}), with a total space usage of $3(n+U)+o(n+U)$ bits. The total space is then close to the $3n\log U$ bits needed to represent $G$ in plain form, and it can be even less if we use compressed wavelet trees to represent the columns.

\subsection{Moving between tables}
We can move from a table to the next one using $C_j$ and $A_j$, for $j \in \{\ts, \tp, \to\}$, using the function $F_j: [1\dd n] \rightarrow [1\dd n]$, defined as follows:
\begin{equation} \label{eq:lf}
F_j(i) ~:=~ A_j[c] + \rank_c(C_j,i),
\end{equation}
where $c=C_j[i]$. Function $F_\to$ maps a position in table $T_{\textsc{spo}}$, using $A_\to$ and $C_\to$, to the corresponding one in $T_{\textsc{osp}}$. In Fig.~\ref{fig:ring}, the straight dashed line maps from $C_\to[7]=3$ to the position of that $3$ in $T_{\textsc{osp}}$, with $F_\to[7]=A_\to[3]+\rank_3(C_\to,7)=6+2=8$. Similarly, $F_\tp$ maps from $T_{\textsc{osp}}$ to $T_{\textsc{pos}}$, and $F_\ts$ maps from $T_{\textsc{pos}}$ back to $T_{\textsc{spo}}$. The function takes $O(\log U)$ time. We can also move in the opposite direction, with the same time complexity, by computing the inverse function of $F_j$ from Eq.~(\ref{eq:lf}): let $c$ satisfy $A_j[c] < i' \le A_j[c+1]$, then
\begin{equation} \label{eq:invlf}
F_j^{-1}(i') := \select_c(C_j,i'-A_j[c]).
\end{equation}

Every node $v$ in the trie of \textsc{spo} corresponds to a range of rows in $T_{\textsc{spo}}[s \dd e]$ (i.e., a range in $C_\to$): if $v$ is the root, the range is $C_\to[s\dd e]=[1\dd n]$.
If $v$ is in the first level and corresponds to the subject $\ts = x$, the range $C_\to[s\dd e]$ is that of all triples starting with $x$. If $v$ is in the second level and corresponds to $(\ts,\tp) = (x,y)$, then $C_\to[s\dd e]$ corresponds to the triples starting with $xy$. A leaf trie node denoting the triple  $(\ts,\tp,\to) = (x,y,z)$ corresponds to a single position in $T_{\textsc{spo}}$ containing $xyz$ (i.e., a cell in $C_\to)$. The same holds, analogously, for tables $T_{\textsc{osp}}$ (column $C_\tp$) and $T_{\textsc{pos}}$ (column $C_\ts)$. The other three tries are also implicitly represented by the tables. Consider the trie for \textsc{pso}. A first-level node for $\textsc{p}=x$ corresponds to a range of rows in $T_{\textsc{pos}}$ (i.e., in $C_\ts$), a second-level node representing $(\tp,\ts) = (x,y)$ corresponds to a range of rows in $T_{\textsc{spo}}$ (i.e., in $C_\to)$, and so on.

Along the search, each triple pattern will have a bound subset of attributes $\{ \textsc{s}, \textsc{p}, \textsc{o}\}$, which always matches a prefix $X$ of either \textsc{spo}, \textsc{osp}, or \textsc{pos}, the three tries we represent via columns $C_\to$, $C_\tp$, and $C_\ts$. As explained, every concrete value for a prefix $X$ corresponds to a range in some column. As we progress, the set $X$ expands and we may have to switch from one column to another. For example,  
given the range $C_\to[s\dd e]$ (i.e., $T_{\textsc{spo}}[s\dd e]$) of the triples sharing a prefix $X$ of $\textsc{spo}$, we obtain the range $C_\tp[s'\dd e']$ (i.e., $T_{\textsc{osp}}[s'\dd e']$) of the triples sharing prefix $cX$ of \textsc{osp} with 
\begin{eqnarray}
s' & := & A_{\textrm{o}}[c] + \rank_c(C_{\textrm{o}},s-1)+1, \nonumber \\
e' & := & A_{\textrm{o}}[c] + \rank_c(C_{\textrm{o}},e). 
\label{eq:range_mapping}
\end{eqnarray}
This is called a {\em backward step}. Fig.~\ref{fig:ring} shows how we descend from the first-level node $8$ in $T_\textsc{pos}$ (represented by $C_\ts[5\dd 9]$) to its child with value $6$ (represented by $C_\to[5\dd 9]$), and from there to its child with value $3$ (represented by $C_\tp[8\dd 8]$). An analogous {\em forward step} extends $X$ to $Xc$, in this case restricting the range $C_\to[s\dd e]$ to a smaller range $C_\to[s'\dd e']$ in the same column; see the original article \cite{AHNRRSsigmod21,AGHNRRStods24} for details.
 
\subsection{Constants in triple patterns} \label{sec:ring-constants}

When \LTJ starts, we find a range in some suitable column $C_*$ for the constants of each triple pattern $t_i$. We choose a table $T_*$ ($T_{\textsc{spo}}$, $T_{\textsc{osp}}$, or $T_{\textsc{pos}}$) whose attribute order is prefixed by the constant attributes in $t_i$, and find the range corresponding to the constant prefix $X$ in the column that represents $T_*$. For example, if only the attribute \to\ is the constant, we start from $T_{\textsc{spo}}[1\dd n]$ and apply Eq.~(\ref{eq:range_mapping}) to end with some $T_{\textsc{osp}}[s\dd e]$; if $\tp$ and $\ts$ are the constants, we start from $T_{\textsc{osp}}[1\dd n]$ and apply (the analogous of) Eq.~(\ref{eq:range_mapping}) twice to end with some $T_{\textsc{spo}}[s\dd e]$. The total initialization time is $O(\log U)$ per triple pattern.


\subsection{Supporting leaps} \label{sec:ring-leap}
The remaining piece to support \LTJ is function $\leap{t_i',c}$ (Section~\ref{subsec:ltj}), where $t_i'$ is either a triple pattern $t_i$ from $Q$, or one of its progressively bound versions $\mu(t_i)$. This finds the smallest child of the current node of $t_i'$ with value $c_x \ge c$. In the context of the \ring, this is done differently depending on whether or not the variable appears to the left of the current prefix matched. If it does, for example we are binding $\to$ and our range is $T_{\textsc{spo}}[s\dd e]$, then we use $\rangenext(T_{\textsc{spo}},s,e,c)$ (Section~\ref{subsec:wt}) to find the appropriate value of $c_x$, and if we decide to assign that value to $\to$ we use Eq.~(\ref{eq:range_mapping}). A more difficult case arises when the desired variable is not to the left, as if binding $\tp$ in $T_{\textsc{spo}}[s\dd e]$. This only happens when we have bound just one position so far, so we start from the range 
$T_{\textsc{pos}}[A_\tp[c]+1 \dd n]$, rework Eq.~(\ref{eq:range_mapping}) for the current value of $\ts$, and finally use Eq.~(\ref{eq:invlf}) to obtain the desired value $c_x$. In all cases, $\leap{}$ takes $O(\log U)$ time and the \ring solves queries in wco time $O(Q^* \log U)$.

\section{RDFCSA: LTJ on a compressed suffix array}\label{sec:rdfcsa} 

We now present a new data structure, which roughly doubles the space of the \ring in exchange for being potentially faster.
The \rdfcsa \cite{BCdBFNsupe22} was designed as a compact representation for labeled graphs that can be queried by single triple patterns and binary joins. It predates the \ring and shares with it the model of viewing the graph triples as cyclic strings of length 3 (in \textsc{spo} order). This set of strings is indexed and compactly represented with a compressed suffix array (\csa, see next). 
The \csa on the cyclic strings suffices to solve the original \rdfcsa queries, but in order
to support the \LTJ algorithm, the \rdfcsa lacks bidirectionality, that is, unlike the \ring, it cannot support $\leap{}$ on variables to the left and to the right of the already bound positions. 

We now extend the \rdfcsa to support \LTJ by storing two {\csa}s, 
one for the \textsc{spo} order, and another for the \textsc{ops} order, and adding them the support for $\leap{}$, in one direction. We expect this implementation to be faster than that of the \ring (Section~\ref{sec:ring-leap}) because the \csa is in practice more efficient than the wavelet tree for this problem, even if both algorithms take logarithmic time.

\subsection{Compressed Suffix Array} \label{sec:csa}
\no{\pending{Required for RDFCSA.: Fari: Lo pongo despues que los bitmaps, pues necesito rank aqui
Contar $\Psi$ \cite{GV00} y el bitvector $D$ }\\
}

Given a string $S[1\dd n]$ of symbols drawn from an alphabet $\Sigma = \{1,\ldots, \sigma\}$ (except the special symbol $S[n] = \$$, which is lexicographically smaller than all symbols in $\Sigma$), the suffix array $A$ of $S$ \cite{MM93} lists all suffix indices $[1\dd n]$ of $S$ in increasing lexicographic order; that is, $S[A[i]\dd n] <  S[A[i+1]\dd n]$ for all $i \in [1\dd n-1]$ .
For example, let $S = abracadabra\$$, then $A =\langle 12,11,8,1,4,6,9,2,5,7,10,3\rangle$. Note that all the occurrences of any given substring pattern $P[1\dd m]$ are pointed from a contiguous range $A[r_s\dd r_e]$ (because they are prefixes of the suffixes $S[A[i] \dd n], i\in A[r_s\dd r_e]$).

The compressed suffix array (\csa) \cite{Sad03} is a compact representation of the suffix array that replaces both $S$ and $A$. It uses a permutation $\Psi[1\dd n]$ such that $\Psi[i] =j$ if $A[j]=A[i]+1$ (or $A[j]=1$ if $A[i]=n)$. Therefore, given a position $p=A[i]$ in $S$, $j=\Psi[i]$ gives the index in $A$ such that $A[j] = p+1$, the next position in $S$.
For the example above we have $\Psi[1\dd n] = \langle 4,1,7,8,9,10,11,12,6,3,2,5 \rangle$. Note that $S[A[7]\dd 12]= bra\$$, $\Psi[7]=3$, and so $S[A[3]\dd 12] = ra\$$. 
The CSA also includes a bitvector $D[1\dd n]$ that sets $D[i]=\bit{1}$ to mark the positions $i$ in $A$ where the first symbol of the suffix pointed to from $A[i]$ changes, that is, $D[i]=\bit{1}$ iff $i=1$ or $S[A[i-1]] < S[A[i]]$. In our example, $D=\langle \bit{1},\bit{1},\bit{0},\bit{0},\bit{0},\bit{0},\bit{1},\bit{0},\bit{1},\bit{1},\bit{1},\bit{0} \rangle$. 
The symbol $c= S[x]$, pointed from $A[i]=x$, can be obtained as $c = \rank_{\bit{1}}(D,i)$. 
Further, $S[x+1]= \rank_{\bit{1}}(D,\Psi[i])$, $S[x+2]= \rank_{\bit{1}}(D,\Psi[\Psi[i]])$, and in general $S[x+k]=\rank_{\bit{1}}(D,\Psi^k[i])$.

Regarding space, $\Psi$ is composed of at most $\sigma$ increasing sequences, which can be compressed by encoding differences and applying run-length encoding for runs of $+1$ differences. The required space is $nH_k(S)+O(n\log\log\sigma)$ bits for any $k \le \alpha \log_\sigma n$ and constant $\alpha<1$ \cite{NM07}, where $H_k(S) \le \lg\sigma$ is the $k$-th order entropy of $S$. Bitvector $D$ adds just $n+o(n)$ bits. 

\blockcomment{
\no{
\subsection{Suffix Array.}
A Suffix Array  is a sorted structure that contains all suffixes of a string. Let $T[1\dd n]$ be a string on an ordered alphabet $\Sigma$. Each symbol $T[i] \in \Sigma$ except T[n] = $\$$. The suffix array $A$ of $T$ stores all suffix indexes of $T$ in increasing lexicographic order. For example let $T = abracadabra\$$, then $A =<12, 11, 8, 1, 4, 6, 9, 2, 5, 7, 10, 3>$. The suffix array is used to locate every occurrence of a substring pattern in a certain range $A[r_s\dd r_e]$. 
}

\begin{figure}[t]
\begin{center}
\includegraphics[scale=0.4]{figures/suffix_array.png}
\caption{Suffix Array of T = abracadabra\$.}
\label{fig:suffix_array_example}
\end{center}
\end{figure}
}

\subsection{Indexing the data} The \rdfcsa requires a particular mapping from $\dom(G)$ to integers. Different alphabets $[1\dd n_\ts]$, $[1\dd n_\tp]$, and $[1\dd n_\to]$ must be considered, respectively, for subjects, predicates, and objects. From them, the first $n_{\textsc{so}}$ symbols in the alphabets of subjects and objects are constants that could occur both as subjects and objects in a triple. It then holds $|U| = n_\tp+ n_\ts + n_\to -n_{\textsc{so}}$.

\begin{figure*}[t]
\centering
\includegraphics[width=1.0\textwidth]{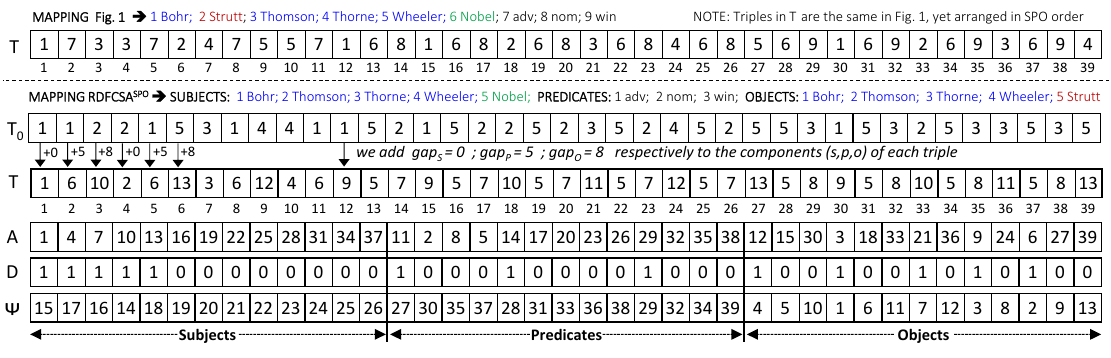}
\caption{Structures involved in the construction of \rdfcsaspo (i.e., $D$ and $\Psi$) for the graph in Fig.~\ref{fig:graph_mapping_orders}.}
\label{fig:rdfcsa-wco}
\end{figure*}

Considering a sequence of $n$ triples sorted in \textsc{spo} order, the \rdfcsa creates a unique sequence of integers $T[1\dd 3n]$ where, for each triple $(s,p,o) \in G$, the string $\langle s'\!,p'\!,o'\rangle = \langle s+gap_\ts, p+gap_\tp,o + gap_\to \rangle$ is appended to $T$. The offsets $(gap_\ts,gap_\tp,gap_\to) = (0,n_\ts,n_\ts\!+n_\tp)$ enforce disjoint identifiers for subjects, predicates, and objects, and ensure $s' < p' <o'$. Then, a \csa is built on $T$. %
Because of the offsets, there are three regions in the suffix array $A$ (and $D$), $A[1\dd n]$, $A[n+1\dd 2n]$, $A[2n+1\dd 3n]$, with entries pointing respectively to the subjects, predicates, and objects in $T$. Consequently, $\Psi[1\dd n]$ contains only values within $[n+1,2n]$, whereas the values in $\Psi[n+1\dd 2n]$ and $\Psi[2n+1\dd 3n]$ are within $[2n+1,3n]$ and $[1,n]$, respectively.
Finally, $\Psi$ is modified to make it cycle on the triples, that is, we enforce $\Psi[\Psi[\Psi[i]]] = i$. This is easily done by decrementing the values in $\Psi[2n+1\dd3n]$, except that $\Psi[i]=1$ is converted to $\Psi[i] = n$.
To reduce space, $\Psi$ is represented as the sequence $\Psi[i]-\Psi[i-1]$, using Huffman and run-length encoding on those gaps. Access in time $O(t_\Psi)$ to any $\Psi[i]$ value is supported by sampling values $\Psi[1+k\cdot t_{\Psi}], k\geq 0$, which requires $O((n/t_\Psi)\log n)$ additional bits on top of the compressed sequence (we will asume $t_\Psi \in O(1)$). Bitvector $D$ takes $3n+o(n)$ further bits.
Fig.~\ref{fig:rdfcsa-wco}  shows an example.


Analogously, we create a second \rdfcsa considering triples sorted in \textsc{ops} order, and with offsets $(gap_\ts,gap_\tp,gap_\to) = (n_\to\!+n_\tp,n_\to,0)$. 
%
We refer to our two \rdfcsa structures as \rdfcsaspo and \rdfcsaops. In either of them, the triple content pointed at position $i$ is retrieved in $O(1)$ time by extracting $\rank_{\bit{1}}(D,i)$, $\rank_{\bit{1}}(D,\Psi[i])$, and $\rank_{\bit{1}}(D,\Psi[\Psi[i]])$), permuting them to order \textsc{spo}, and subtracting the corresponding $gap$ values.
We now show how we carry out the critical processes of \LTJ with these structures.

\subsection{Constants in triple patterns} \label{sec:rdfcsa-constants}
We use the text searching capabilities of the \rdfcsa to find a suffix array interval corresponding to all the triples that match the constants of a given triple pattern. The subsequent variable intersection process then starts from those intervals, which correspond to trie nodes in \LTJ, as with the \ring. 
Recall that, before finding any constant in the \rdfcsa, it must be mapped by adding the corresponding $gap$. We use two operations:
\begin{itemize}
    \item $[l,r] := \rangeCsa(c)$. For a given constant $c$ we obtain the suffix array range $A[l,r]$ of the (cyclic) triples starting with $c$, with $l\leftarrow \select_{\bit{1}}(D,c)$ and $r\leftarrow \select_{\bit{1}}(D,c+1) -1$. This takes constant time. Note that, since $c$ can be a subject, a predicate, or an object, and those identifiers have disjoint suffix array areas in the \rdfcsa, this operation lets us select all the triples with a given subject, a given predicate, or a given object.
    \item $[l,r] := \downCsa(l_c,r_c, d)$. Given a suffix array range $[l_c,r_c] \subseteq \rangeCsa(c)$, so the triples in $A[l_c,r_c]$ start with constant $c$, this operation finds the subrange $[l,r] \subseteq [l_c,r_c]$ of those triples where the $c$ is followed by constant $d$. This is equivalent to stating that  $\forall i\in [l,r], \Psi[i] \in \rangeCsa(d)$. 
    Since $\Psi$ is increasing inside $\rangeCsa(c)$, we can binary search for the first (last) position $l$ ($r$) in $[l_c,r_c]$ such that $\Psi[l]$ ($\Psi[r]$) falls into $\rangeCsa(d)$, in $O(\log n)$ time. 
\end{itemize}

If a triple pattern $t$ has no constants, its range is $[1\dd 3n]$ in both \rdfcsaspo and \rdfcsaops. 
If it has a single constant $c$, then its range in both is $\rangeCsa(c)$ (the mapping of $c$ using $gap$ and the resulting range differs in both $\rdfcsa$s). Which $\rdfcsa$ will be used depends on whether the next variable to eliminate is to the left or to the right of $c$. Therefore, some triple patterns will have a range in \rdfcsaspo and others in \rdfcsaops.

If $t$ has two constants, we can use either $\rdfcsa$ because the next variable to eliminate will be both to the left and to the right of the bound positions. Say we choose \rdfcsaspo. Let $cd$ be the two consecutive constants in \textsc{spo} order (i.e., $\textsc{sp} = cd$, $\textsc{po} = cd$, or $\textsc{os} = cd$). We thus compute $[l_c,r_c] := \rangeCsa(c)$ and then the desired range is $[l_d,r_d] := \downCsa(l_c,r_c,d)$.

\subsection{Supporting leaps}
To support the operation $\leap{}$ we define new primitives: 
\begin{itemize}
    \item $l' := \findTargetPsiCsa(l,r,t_l,t_r)$. Given a range $[l\dd r]$ where $\Psi$ is increasing, it returns the smallest $l'\in[l\dd r]$ such that $\Psi[l'] \in[t_l\dd t_r]$, and $0$ if there is none. It proceeds as for $\downCsa$, yet it needs only one binary search in $[l,r]$.
    \item $l' := \findTargetPsiPsiCsa(l,r,t_l,t_r)$. Given a range $[l\dd r]$ where $\Psi$ is increasing, it returns the smallest $l'\in[l\dd r]$ such that $\Psi[\Psi[l']] \in[t_l\dd t_r]$, or zero if there is none.
    \item $L := \limitV(v)$. It returns the highest offset in $D$ for any constant of the same type as variable $v$. For example, if $v$ is a subject this is $n$ in \rdfcsaspo and $3n$ in \rdfcsaops.
\end{itemize}

Recall that $\leap{t_i',c}$ returns the first constant $c_x  \geq c$ where a given variable $x$ has occurrences in $t_i'$, where $t_i'$ is either a triple pattern $t_i$ from $Q$, or one of its progressively bound versions $\mu(t_i)$. The way we solve $\leap{}$ depends on where the constant(s) and the variable $x$ appear in $t_i'$. 

If there are no constants in $t_i'$, the answer is simply $c_x := c$, because our mapping makes all the symbols appear in $T$. 

If there is only one constant $d$ in $t_i'$, then we have a range $[l,r]$ for $t_i'$ in both \rdfcsaspo and \rdfcsaops. 
If $x$ follows $d$ in \textsc{spo} order, we use \rdfcsaspo, otherwise we use \rdfcsaops.
We first compute $l' := \findTargetPsiCsa(l,r, l_c ,\limitV(x))$ where $l_c = \select_{\bit{1}}(D,c)$. 
Then, if $l' \neq 0$ we return $c_x := \rank_{\bit{1}}(D,\Psi[l'])$, otherwise we return $c_x :=\, \perp$.

Finally, if $t_i'$ contains two constants $d$ and $d'$, they have a range $[l,r]$ for $dd'$ or $d'd$ in either \rdfcsaspo or \rdfcsaops, so we complete the search in the corresponding structure. In this case, we first compute $l' := \findTargetPsiPsiCsa(l,r, l_c ,\limitV(x))$, where $l_c = \select_{\bit{1}}(D,c)$. 
Then, if $l' \neq 0$ we return $c_x:= \rank_{\bit{1}}(D,\Psi[\Psi[l']]$), and $c_x:=\,\perp$ otherwise.
As an example, recall that if we have a triple pattern $(x,y,z)$ with no constants, the initial range in \rdfcsa is $\mathcal{R}:=[1,3n]$. If we bind $y:=\mathsf{win}$ (i.e., the 3rd predicate, which is mapped to id $\mathbf{8} = 3+gap_\tp$), we update $\mathcal{R}:=\rangeCsa(\mathbf{8}) = [23,25]$ in \rdfcsaspo (and also in \rdfcsaops).
Now, since $\mathsf{Wheeler}$ is the 4th object and maps into id $\mathbf{12}=gap_\to +4$,
if we call $\leap{(x,\mathbf{8},z}, \mathbf{12})$, since $\mathsf{Wheeler}$ follows $\mathsf{win}$ in \textsc{s\underline{po}} order, we must use \rdfcsaspo.
We first compute $l' := \findTargetPsiCsa(23,26, \select_\bit{1}(D,\mathbf{12}) , 39) = 26$, and then solve $\leap{(x,\mathbf{8},z), \mathbf{12}} = \rank_\bit{1}(D,\Psi[26]) = \rank_\bit{1}(D,39) = \mathbf{13}$. Therefore,  $\leap{}$ returns object $5=\textbf{13}-gap_\to$ which corresponds to $\mathsf{Strutt}$, that is, the first object after $\mathsf{Wheeler}$ reached from the current range $\mathcal{R}$.

\no{
\section{CompactLTJ: LTJ on compact tries} \label{sec:compactLTJ}

Our second space/time tradeoff, which roughly doubles the space of \rdfcsa, consists in implementing all the 6 tries needed by \LTJ, yet using succinct representations.

\subsection{Trie representation} \label{sec:louds}

The Level-Order Unary Degree Sequence (LOUDS) \cite{Jac89} is a representation of $n$-node tree topologies using just $2n+o(n)$ bits. It is obtained by traversing the tree levelwise (with each level traversed left to right). We append the {\em encoding} $\bit{0}^d\bit{1}$ of each traversed node to a bit sequence $T$, where $d$ is the number of children of the node. The final sequence $T$ represents the tree using two bits per node: a $\bit{0}$ in the encoding of its parent and a $\bit{1}$ to terminate its own encoding. A bitvector representation of $T$ then needs $2n+o(n)$ bits, and allows navigating the tree in constant time.

Our trie topologies are particular in that all the leaves have the same depth, 3. Therefore, every internal node at depths 0--2 have children, and thus we can reduce their encoding to $\bit{0}^{d-1}\bit{1}$. The leaves need not be encoded, which further saves space: we spend exactly {\em one bit} per trie edge, that is, $n+o(n)$ bits for a trie of $n$ nodes, halving the original space \cite{Jac89}.

Our encoding also simplifies the traversal compared to the original LOUDS \cite{Jac89}. We will use the position preceding the encoding of a node as its trie identifier $v \ge 0$. The identifier of the $i$th child of $v$, for $i \ge 1$, is
$\child(v,i) = \select_{\bit{1}}(T, v+i)$. The formula works because $T$ simultaneously enumerates, in levelwise order, the trie edges (one bit per edge) and their target nodes (one $\bit{1}$-terminated encoding per node, leaves omitted). The number of children of $v$ can also be computed in $O(1)$ time as 
$\degree (v) = \selectnext_{\bit{1}}(T, v+1)-v$.

The edge labels are stored in a compact array $L$, each label using $\lceil \lg{U} \rceil$ bits. The labels in $L$ are deployed in the same levelwise order of the edges $T$, so the labels corresponding to the children of node $v$ are all consecutive, in $L[v+1 \dd v+\degree(v)]$. This allows implementing $\leap{}$ efficiently using exponential search from the current position.




\subsection{Indices}

Our main index, \compactLTJ, is exactly as described above, comprising $T$ and $L$. For the trie \textsc{pso} in Fig.~\ref{fig:graph_mapping_orders}, it would store
\begin{eqnarray*}
 T & = & \bit{001 ~ 000111 ~ 1111000010001} \\
 L & = & \mathsf{789 ~ 134566 ~ 3251123451234}
\end{eqnarray*}
where, for example, the second ($i=2$) child of the root ($v=0$) descends by $L[0+2]=\mathsf{8}$ and leads to $u=\child(0,2)=\select_\bit{1}(T,0+2)=7$. The encoding of $u=7$ is at $T[u+1\dd u+\degree(u)] = T[8 \dd 8] = \bit{1}$. Its only child, by $L[7+1] = \mathsf{6}$, leads to $w=\child(7,1)=\select_\bit{1}(T,7+1)=13$. Node $w=13$ then has $\degree(13)=\selectnext(14)-13=5$ children, at $L[14\dd 18] = \mathsf{12345}$.

We also introduce a version called \unCompactLTJ, which is a minimal non-compact trie representation. The \unCompactLTJ index stores an array $P[0\dd]$ of {\em pointers}, one per internal node, deployed in the same order of LOUDS. Pointers are positions in the array using $\lceil \lg n \rceil$ bits. Each internal node $v$ stores in $P[v]$ a pointer to its first child, knowing that the others are consecutive. Its number of children is simply $P[v+1]-P[v]$. Its array $L$ of edge labels is identical to that of $\compactLTJ$. For our example above we have $P= \langle 1,4,8,9,10,11,12,13,14,19,23\rangle$ ($23$ is a terminator). 

In exchange for nearly doubling the space of \compactLTJ, \unCompactLTJ has explicit pointers just like classical data structures, so it does not spend time in computing addresses. As we show in the experiments, \unCompactLTJ still uses half the space of Jena LTJ~\cite{HRRSiswc19}, a classic index that supports LTJ using the six tries (implemented as B+-trees).
}



\section{URing: A unidirectional Ring} \label{sec:uring}

Bidirectionality is the key to using just one \ring to index the $3!=6$ orders required by \LTJ. The \rdfcsa, instead, requires two copies of the index, thereby roughly doubling the space. We now explore the fact that the wavelet tree representation of the ring columns $C_*$ supports an intersection algorithm ($\rangeint$, Section~\ref{subsec:wt}) that is likely faster than the one implemented in \LTJ, which is based on the primitive $\leap{}$. 

When we eliminate a new variable $x$, every triple pattern $t_i$ where it appears is represented by a range in some column, $C_*[l_i\dd r_i]$. Assume $x$ appears to the left of the positions already bound. The desired constants $c_x$ for $x$ are the values that appear in all those ranges $C_*[l_i,\dd r_i]$. We find them by running $\rangeint$ on all those ranges in order to obtain, one by one, the desired values $c_x$ (the algorithm runs even if the ranges are in different sequences $C_*$). We then add each such binding $(x := c_x)$ to the mapping $\mu$ and recurse on that branch.  

The problem with using that intersection algorithm is that it works only if the variable to eliminate is to the left of the current ranges, and therefore, analogously to the \rdfcsa, we have a unidirectional index. Just as for \rdfcsa, we must then have two indices, $\ring^{\textsc{spo}}$ and $\ring^{\textsc{ops}}$ to ensure that we always 
have a range that can be extended to the left. The algorithm proceeds exactly as the \rdfcsa over those two copies, except that the intersection algorithm of \LTJ is replaced by the custom 
algorithm $\rangeint$. An additional benefit is that going rightwards in the binding is somewhat more expensive on the \ring than going leftwards, and this new variant goes always leftwards.

%
%

\section{Improved Variable Elimination Orders} 

Our second contribution is the study of improved VEOs in the context of compact indices for LTJ, which deviate from the VEO defined in Section~\ref{sec:veo}. The first improvement is the use of {\em adaptive} VEOs; the second is on how to efficiently compute (or approximate) $w_{ij}$ in our compact index representations.

\subsection{Adaptive VEOs} \label{sec:adaptive}

In previous work using the VEO described in Section~\ref{sec:veo}, the VEO is fixed before running \LTJ. The selectivity of each variable $x_j$ is estimated beforehand, by assuming it will be the {\em first} variable to eliminate. In this case, Eq.~(\ref{eq:wj}) takes the minimum of the number of children in all the trie nodes we must intersect, as an estimation of the size of the resulting intersection. The estimation is much looser on the variables that will be eliminated later, because the children to intersect can differ a lot for each value of $x_j$.

We then consider an {\em adaptive} version of the heuristic: we use the described technique to determine only the {\em first} variable to eliminate. Say we choose $x_j$. Then, for each distinct binding $x_j := c$, the corresponding branch of \LTJ will run the VEO algorithm again in order to determine the second variable to eliminate, now considering that $x_j$ has been replaced by $c$ in all the triples $t_i$ where it appears. This should produce a much more accurate estimation of the intersection sizes.

In the adaptive setting, we do not check anymore that the new variable shares a triple with a previously eliminated one; this aimed to capture the fact that those triples would be more selective when some of their positions were bound, but now we know exactly the size of those progressively bound triples. The lonely variables are still processed at the end.

Observe that, in the adaptive case, there is not anymore a single VEO; each different branch can have one. While this technique does not require any extra space, it could incur in an extra cost to repeatedly recompute the VEO (in fact, only its first variable) for each binding. The cost to compute Eq.~(\ref{eq:wj}), on each of our compact index representations, becomes then of paramount importance for adaptive VEOs.

\subsection{Computing the VEO predictors} \label{sec:vring}

The \ring cannot efficiently compute $w_{ij}$ as described in Section~\ref{sec:veo}, because it does not know the number of children of the node $v$: the \ring only has the range $C_*[l\dd r]$ corresponding to $v$, which results from resolving the constants in $t_i$; recall Section~\ref{sec:ring-constants}. The size $r-l+1$ of the range was then used as a reasonable estimation of $w_{ij}$ \cite{AHNRRSsigmod21,AGHNRRStods24}. In the case $t_i$ has one constant and $x_j$ is to its right, we can also use $r-l+1$ to estimate $w_{ij}$, because the range size would be the same in both directions. We do the same when implementing the \rdfcsa.

While the use of $r-l+1$ as a predictor of the true weight $w_{ij}$ can be seen as an approximation, it can be argued to be a {\em better} predictor of the difficulty of binding $x_j$ in $t_i$. Note that $r-l+1$ is the number of {\em leaf descendants} of the current node $v$ of the trie $\tau_i$, whose children are the bindings of $x_j$ in $t_i$. The number of descendants may be a more accurate estimation of the {\em total} work that is ahead if we bind $x_j$ in $t_i$, as opposed to the children, which yield the number of distinct values $x_j$ will take without looking further.

\no{The \compactLTJ index uses the original estimator based on the number of children of $v$, which is easily computed in constant time as  $w_{ij} = \degree(v)$. We now define an alternative version, \compactLTJstar, which computes $w_{ij} = r-l+1$, the number of leaf descendants of $v$. This is computed as $w_{i,j}=n$ if $v$ is in the first level, and $w_{i,j}=\degree(v)$ if $v$ is in the third level. For the second level, we compute in
constant time $w_{ij} = \child(v+\degree(v),1)-\child(v,1)$.
}

By using (additional) wavelet trees, we can also compute $w_{ij}$ as the number of children of $v$ on the \ring, in $O(\log n)$ time \cite{GKNP13}. What we need is to count the number of {\em different} symbols in the range $C_*[l\dd r]$. Let $M$ be such that $M[i]$ is the largest value $i' < i$ such that $C_*[i] = C_*[i']$, or $0$ if there is no such $i'$. This implies that $C_*[q]$ is the {\em first} occurrence of a symbol in $C_*[l \dd r]$ iff $l \leq q \leq r$ and $M[q] < l$. 
We can then count the number of first occurrences of symbols in $C_*[l\dd r]$ by counting the number of values less than $l$ in $M[l\dd r]$. This is accomplished by the $\rangecount$ function (Section~\ref{subsec:wt}) if we have $M$ represented as a wavelet tree. 

Note that the use of $M$ requires that the new variable $x_j$ is to the left of the constants in $t_i$; therefore we need one sequence $M$ per column $C_*$ in both the (actual) \ring for the order \textsc{spo} and the (virtual) \ring for the order \textsc{ops}. So, even if we use just one bidirectional \ring, we must add two sequences $M$ (thus tripling the space).  

\subsection{Refining VEO predictors} \label{sec:veo-inters}

The value $w_j$ obtained by using the VEO predictor where the estimator is the size of the range $C_*[l\dd r]$ corresponds to the maximum number of triples that can participate in the intersection. Therefore, it is an upper bound to the size of the intersection of the set of triples. 

We can obtain a better approximation of the intersection size by splitting the values of $C_*[l\dd r]$ into disjoint subsets of the alphabet and applying Eq.~(\ref{eq:wj}) to each. The sum of the weights of the subsets is a more refined approximation to the intersection. We can then refine the heursitic of Eq.~(\ref{eq:wj}) as
\begin{equation} \label{eq:wj-inter}
w_j = \sum_{\gamma \subset [1, \sigma]}{\min \{w^\gamma_{ij} ~|~ x_j \textrm{ appears in triple } t_i\}},
\end{equation}
where $w^\gamma_{ij}$ is the weight of $x_j$ in $t_i$ for the partition $\gamma$ of the alphabet. The partitions are disjoint and their union is $[1, \sigma]$. 

The \textit{ring} can exploit the wavelet trees to easily compute $w^\gamma_{ij}$. Let us consider a variable $x_j$ that appears in a triple pattern $t_i$, with the possible values of $x_j$ in the range $C_*[l\dd r]$. Our estimation algorithm starts in the range $B_{1,\sigma}[l\dd r]$ of the root. By mapping $[l \dd r]$ to its left child ($B_{1,m}$), where $m$ is $\lfloor(\sigma+1)\rfloor/2$, we retrieve the range $B_{1,m}[l'\dd r']$ of the symbols from $C_*[l\dd r]$ that belong to the first half of the alphabet $[1, m]$. 
That range is computed in $O(1)$ time
as $[rank_0(B_{1,\sigma},l-1)+1 \dd rank_0(B_{1,\sigma},r)]$. The length of this range is equivalent to $w^{[1,m]}_{ij}$. By using $rank_1$ instead of $rank_0$ we compute the range of $[l \dd r]$ in the right child ($B_{m+1,\sigma}$), whose length is $w^{[m+1,\sigma]}_{ij}$. Since $w^{[1,m]}_{ij}+w^{[m+1,\sigma]}_{ij}=w_{ij}$ for every $i,j$, $(\min_i w^{[1,m]}_{ij})+(\min_i w^{[m+1,\sigma]}_{ij}) \le \min_i w_{ij}$ is a tighter upper bound on the size of the intersection.

In that procedure we obtain two partitions, $[1,m]$ and $[m+1,\sigma]$. If we continue mapping the current ranges in $B_{1,m}$ and $B_{m+1,\sigma}$ to their children, each previous partition splits into two other halves. Therefore, by repeating those steps $k$ levels, we get up to $2^k$ partitions. The length of each range in a node of the $k$-th level matches a weight $w^\gamma_{ij}$. As $k$ increases, the approximation of the heuristic improves (indeed, if we reach $k=\lg \sigma$ we obtain the actual size of the intersection, at cost $O(\sigma)$). 

\begin{table*}[t]
    \centering
    \begin{tabular}{l@{~~~}c@{~~~~}c@{~~~}c@{~~~}c@{~~~}r}
    \toprule
      $k$ & $\gamma$ & $w^\gamma_{ij}$   & $w^\gamma_{i'j}$ & $\min$ & $w_j$ \\
      \midrule
       0  & $[1,6]$  & 4 & 4 & 4 & 4 \\   
       \midrule
        \multirow{2}{*}{1}  & $[1,3]$  & 2 & 0 & 0 & 
        \multirow{2}{*}{2} \\
                  & $[4,6]$  & 2 & 4 & 2 & \\
        \midrule
        \multirow{4}{*}{2}  & $[1,2]$  & 1 & 0 & 0 & \multirow{4}{*}{0} \\
          & $[3,3]$  & 1 & 0 & 0 & \\
          & $[4,5]$  & 2 & 0 & 0 & \\
          & $[6,6]$  & 0 & 4 & 0 & \\
      \bottomrule
    \end{tabular}
    \caption{The partitions $\gamma$ and weights obtained in the sequence of Fig.~\ref{fig:wt} with the ranges $[1,4]$ and $[5,8]$ of two triple patterns $t_i$ and $t_{i'}$, respectively. Column $k$ indicates the number of levels to traverse in the wavelet tree.}
    \label{tab:example}
\end{table*}

For example,
consider two triples, $t_i$ and $t_{i'}$, whose defined ranges are $[1 \dd 4]$ and $[5 \dd 8]$, respectively, in the sequence of Fig.~\ref{fig:wt}. Note that their intersection is empty. Since both ranges have length $4$, the weight $w_j$ per Eq.~(\ref{eq:wj}) is $4$. In order to apply Eq.~(\ref{eq:wj-inter}), the algorithm starts at the root $B_{1,6}$ and maps each range to its children. For example, the range $B_{1,6}[1 \dd 4]$ maps to $B_{1,3}[1 \dd 2]$ and $B_{4,6}[1 \dd 2]$, thus $w^{[1,3]}_{ij}=2$ and $w^{[4,6]}_{ij}=2$. In the same way, from $B_{1,6}[5 \dd 8]$ we just reach $B_{4,6}[3 \dd 6]$, so $w^{[1,3]}_{i'j}=0$ and $w^{[4,6]}_{i'j}=4$. Consequently, $w_j = \min(w^{[1,3]}_{ij}, w^{[1,3]}_{i'j}) + \min(w^{[4,6]}_{ij},w^{[4,6]}_{i'j}) = 2$, which is a closer bound to the actual intersection size. By descending one more level from the ranges in $B_{1,3}$ and $B_{4,6}$, we will obtain four partitions and $w_j$ reaches $0$. The partitions and weights obtained for each level are shown in Table~\ref{tab:example}. 

\medskip

Note that we are assuming that the range in the \textit{ring} contains the values of the variable to bind $x_j$. However, it is possible that in $t_i$ the defined range does not correspond to that variable. Those cases can occur when only one position of $t_i$ is bound. For instance, if the predicate of $t_i$ is bound to a constant $c$, the \textit{ring} defines the range $C_\ts[l\dd r]$ ($T_{\textsc{pos}}$). When the variable $x_j$ is the object of $t_i$, there is no direct access to the values of $x_j$. 

In that case, as in the main algorithm, we compute the size of each partition $\gamma$ in $C_\to[1\dd n]$ ($T_{\textsc{spo}}$), without any interval restriction. With this approach, $w^\gamma_{ij}$ is equivalent to the number of triples whose object belongs to $\gamma$. From those, we just need to count the triples where its predicate has constant $c$. Since those partitions correspond to non-overlapping consecutive intervals in $T_{\textsc{osp}}$, the number of triples whose predicate is $c$ in each alphabet partition $\gamma$ is obtained as $rank_c(C_\tp,e) - rank_c(C_\tp,s)+1$, where $[s,e]$ is the range of $\gamma$ in $T_{\textsc{osp}}$. This range is computed as $s=A_{\textrm{o}}[c_s]+1$ and $e=A_{\textrm{o}}[c_e+1]$,  where $c_s$ and $c_e$ are the first and last symbol, respectively, of each alphabet partition $\gamma$.

\section{Experimental results}

We compare the compact indexing schemes described along the paper and various state-of-the-art alternatives, in terms of space usage and time for evaluating various types of BGPs. 

Our experiments ran on an Intel(R) Xeon(R) CPU E5-2630 at 2.30GHz, with 6 cores, 15 MB cache, and 378 GB RAM.

\subsection{Datasets and queries}

We run two benchmarks over the Wikidata graph~\cite{VrandecicK14}, which we choose for its scale, diversity, prominence, data model (i.e., labeled edges) and real-world query logs~\cite{MalyshevKGGB18,BonifatiMT19}. The graph features $n=958{,}844{,}164$ triples, 
which take 10.7 GB if stored in plain form using 32 bits for the identifiers. 

We consider a real-world query log \cite{MalyshevKGGB18}. In search of challenging examples, we downloaded queries that gave timeouts, and selected queries with a single BGP, obtaining $1{,}295$ unique queries. Those are classified into three categories: (I) $520$ BGPs formed by a single triple pattern, which mostly measure the retrieval performance of the index; (II) $580$ BGPs with more than one triple but only one variable appearing in more than one triple, which measure the performance of joins but do not distinguish good from bad VEOs (as long as the join variable is eliminated first, of course); (III) $195$ complex BGPs, where the performance of different VEOs can be compared.

\no{
A second set of queries is obtained from the Wikidata Graph Pattern Benchmark (WGPB) \cite{HRRSiswc19}, proposed for a subgraph of Wikidata with diverse graph patterns. We run them on the full graph.
%
%
The benchmark provides 17 query patterns of different widths and shapes, including cyclic and acyclic queries, as shown in Fig.~\ref{fig:wikidata-subgraph-query-patterns}. Each pattern is instantiated with 50 queries built using random walks such that the results are non-empty. The benchmark allows us to compare different index schemes and databases for different abstract patterns. All predicates are constant, all subjects and objects are variables, and each variable appears at most once in the same triple pattern. 


\begin{figure}[t]
    \centering
    \input{figures/WikidataGraphPatterns}
    \caption{Query patterns for the Wikidata benchmark.}
    \label{fig:wikidata-subgraph-query-patterns}
\end{figure}
}

All queries are run with a timeout of 10 minutes and a limit of 1000 results (as originally proposed for WGPB~\cite{HRRSiswc19}). This measures the time the systems need to display a reasonable number of results. We also compare the systems without the limit of results, which measures throughput in cases where we need all the results. The space of the indices is measured in bytes per triple (bpt); a plain 32-bit storage requires 12 bpt.

\subsection{Systems compared}
\label{subsec:exp_setup}

Our experiments compare all indexing schemes described:
\begin{itemize}
    \item Two \ring variants (Section~\ref{sec:ring}), \Ring-large and \Ring-small, corresponding to using plain or compressed bitvectors in the wavelet trees, respectively (these are called Ring and C-Ring, respectively, in the original paper \cite{AHNRRSsigmod21,AGHNRRStods24}).
    \item Their corresponding unidirectional versions, which compute intersections using the wavelet tree (Section~\ref{sec:uring}): \URing-large and \URing-small.
    \item Their extension to compute the standard VEO based on number of children (Section~\ref{sec:vring}): 
    \VRing-large, \VRing-small, \VURing-large and \VURing-small.
    \item Their extension to compute the refined estimators for the VEO (Section~\ref{sec:veo-inters}), both in their bidirectional and unidirectional variantes: \IRing-large, \IRing-small, \IURing-large, and \IURing-small. The number of levels that descend those estimators is configured to $3$.
    \item Two \rdfcsa variants (Section~\ref{sec:rdfcsa}): \Rdfcsa-large represents $\Psi$ in plain form; 
    \Rdfcsa-small sets $t_{\Psi}=16$ and uses Huffman and run-length encoding to compress $\Psi$.
    \item All the versions above compute the VEO in traditional (``global VEO'') and in adaptive form (Section~\ref{sec:adaptive}).
\end{itemize}


We also compare various prominent graph database systems:

\begin{itemize}
\item MillenniumDB \cite{VR+23}: A recently developed open-source graph database. 
We use here a specialized version that stores six tries in the form of B+-trees and supports full \LTJ,  with a sophisticated (yet global) VEO. We run MillenniumDB over a RAM disk to avoid using external memory.


\item Jena LTJ \cite{HRRSiswc19}: An implementation of \LTJ on top of Apache Jena TDB. All six different orders on triples are indexed in B+-trees, so the search algorithm is always wco. 
\item RDF-3X \cite{NW10}: Indexes a single table of triples in a compressed clustered B+-tree. The triples are sorted and those in each tree leaf are differentially encoded. 
RDF-3X handles triple patterns by scanning ranges of triples and features a query optimizer using pair-wise joins.
\item Virtuoso \cite{Erling12}: The graph database hosting the public DBpedia endpoint, among others. It provides a column-wise index of quads with an additional graph ($g$) attribute, with two full orders (\textsc{psog}, \textsc{posg}) and three partial indices (\textsc{so}, \textsc{op}, \textsc{gs}) optimized for patterns with constant predicates. 
It supports nested loop joins and hash joins.
%
\item Blazegraph \cite{Blazegraph}: The graph database system hosting the official Wikidata Query Service~\cite{MalyshevKGGB18}. We run the system in triples mode, with B+-trees indexing orders \textsc{spo}, \textsc{pos}, and \textsc{osp}. It supports nested-loop joins and hash joins. 
%

\end{itemize}

\no{We exclude Graphflow \cite{graphflow}, ADOPT \cite{adopt}, and EmptyHeaded \cite{EmptyHeaded} because we have not enough memory to build them.
Section~\ref{sec:beyond-wco} compares them on a smaller graph. }

The code was compiled with $\mathsf{\text{g++}}$ with flags $\mathsf{\text{-std=c++11}}$ and $\mathsf{\text{-O3}}$; some alternatives have extra flags to enable third party libraries. 
Systems are configured per vendor recommendations. 



\subsection{Results}
\label{sec:real-world_benchmark}

\begin{figure}[t]
\begin{center}
    \includegraphics[width=0.6\textwidth]{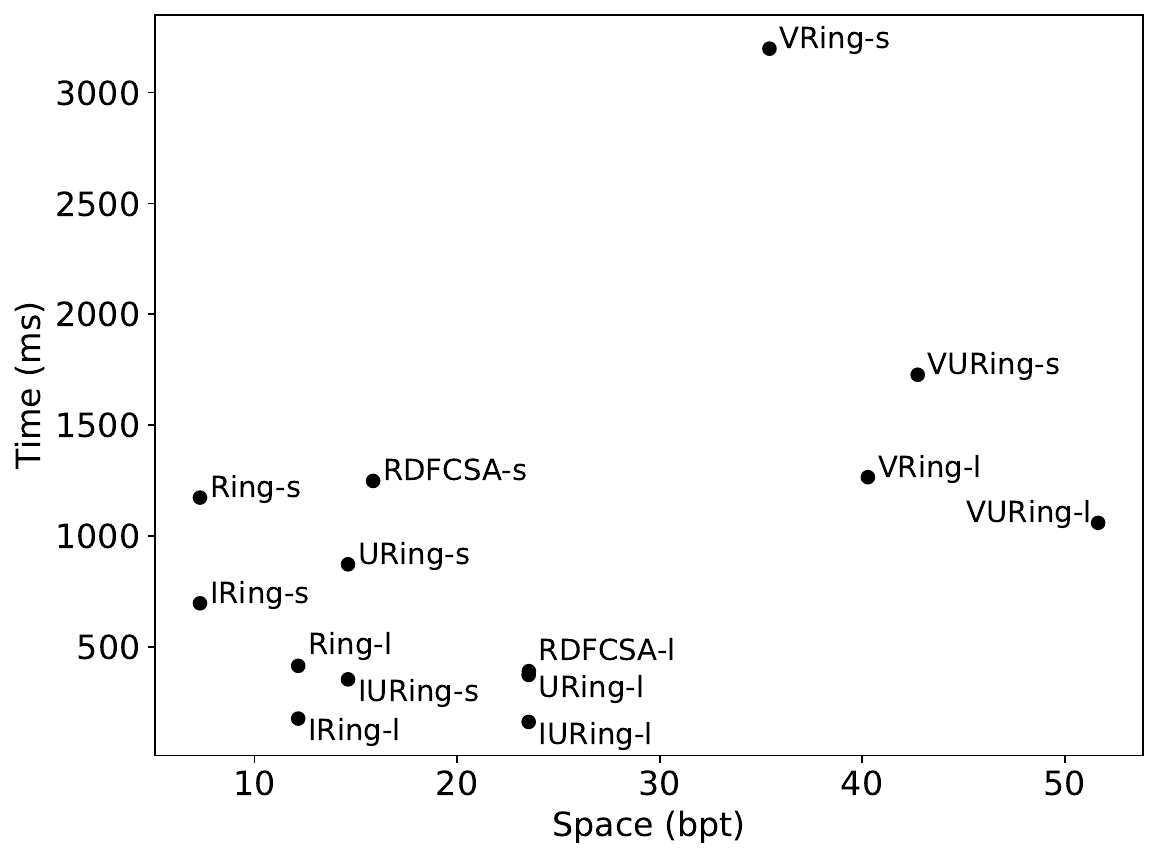}
    \caption{Index space and the averaged query times of the adaptive variants in msec, limiting outputs to 1000 results. Suffixes \textit{s} and \textit{l} mean \textit{small} and \textit{large}, respectively.}
    \label{fig:averages}
\end{center}
\end{figure}

Table~\ref{tab:limit} shows the index space in bytes-per-triple (bpt) and some general statistics on the query times obtained on our benchmark, when we limit the results to 1000. We list all the \ring and \rdfcsa variants first, and then other systems. Fig.~\ref{fig:averages} illustrates the tradeoff between index space and average query time for our adaptive variants.


\begin{table*}[t]
  \centering
     \centering
        \begin{tabular}{lrrrrrrrrr}
          \toprule
            System      & Space  & \multicolumn{2}{c}{Average} & \multicolumn{2}{c}{Median} & \multicolumn{2}{c}{Timeouts}\\
           & (bpt) & Gl & Ad & Gl & Ad & Gl & Ad \\
            \midrule
            \Ring-small               &  7.30  & 3056 & 1173 & 24 & 24 & 5 & 0 \\
            \IRing-small              &  7.30  & 2568 & 696 & 27 & 27 & 4 & 0 \\
            \Ring-large               & 12.15  & 2256 & 414 & 8 & 8 & 3 & 0 \\
            \IRing-large              & 12.15  & 2021 & 176 & 8 & 8 & 3  & 0 \\
            \URing-small                       & 14.61  & 2779 & 872 & 20 & 20 & 4 & 1\\
            \IURing-small                       & 14.61  & 2300 & 353 & 24 & 20 & 3 & 0\\
            \URing-large                       & 23.53  & 1481 & 373 & 8 & 8 & 0 & 0 \\
            \IURing-large                       & 23.53  & 1319 & 161 & 8 & 8 & 0 & 0 \\
            \VRing-small                       & 35.42  & 4594 & 3198 & 24 & 24 & 4 & 3\\
            \VRing-large                       & 40.28  & 3067 & 1265 & 8 & 8 & 5 & 1 \\
            \VURing-small                      & 42.74  & 3467 & 1727 & 20 & 20 & 1 & 2\\
            \VURing-large                      & 51.65  & 2124 & 1059 & 8 & 8 & 0 & 1 \\
            \Rdfcsa-small                      & 15.85  & 2323 & 1248 & 8 & 8 & 1 & 1\\
            \Rdfcsa-large             & 23.54  & 579 & 390 & 2 & 2 & 0 & 0\\
            \midrule
            MillenniumDB          &156.78  & \multicolumn{2}{c}{~~\,96} & \multicolumn{2}{c}{~\,27} & \multicolumn{2}{c}{~\,0} \\
            Jena LTJ       & 168.84  & \multicolumn{2}{c}{1930} & \multicolumn{2}{c}{162} & \multicolumn{2}{c}{~\,1} \\
            Virtuoso                    & 60.07   & \multicolumn{2}{c}{4880} & \multicolumn{2}{c}{~\,50} & \multicolumn{2}{c}{~\,8} \\
            RDF-3X                      & 85.73  & \multicolumn{2}{c}{8230} & \multicolumn{2}{c}{126} & \multicolumn{2}{c}{13} \\
            Blazegraph                  & 90.79  &    \multicolumn{2}{c}{9220}   & \multicolumn{2}{c}{~\,54} & \multicolumn{2}{c}{14}  \\
            \bottomrule
        \end{tabular}

    \caption{Space and query times (in msec) of all the systems, limiting results to 1000, with Gl(obal) and Ad(aptive) VEOs. Timeouts count queries exceeding 10 min.}
    \label{tab:limit}
\end{table*}

\begin{figure}[ht!]
    \centering
    \begin{tabular}{lrrrrrrr}
    \toprule
      System & Space   & \multicolumn{2}{c}{Type I} & \multicolumn{2}{c}{Type II} & \multicolumn{2}{c}{Type III} \\
             &   (bpt)            & Avg & Med & Avg & Med & Avg & Med \\
      \midrule
      \IRing-small   &   7.30 & 12 & 8.0 & 380 & 36 & 3455 & 97  \\
      \IRing-large   &  12.15 & 3.9 & 2.9 & 93 & 11 & 881 & 32 \\
       \IURing-large   &  23.53 & 4.6 & 4.0 & 75 & 12 & 832 & 36 \\
      \Rdfcsa-large &  23.54 & 0.6 & 0.3 & 18 & 2.9 & 2611 & 14  \\
      \bottomrule
      \vspace{0.02cm}
    \end{tabular}
        \centering        
        \includegraphics[width=0.85\textwidth]{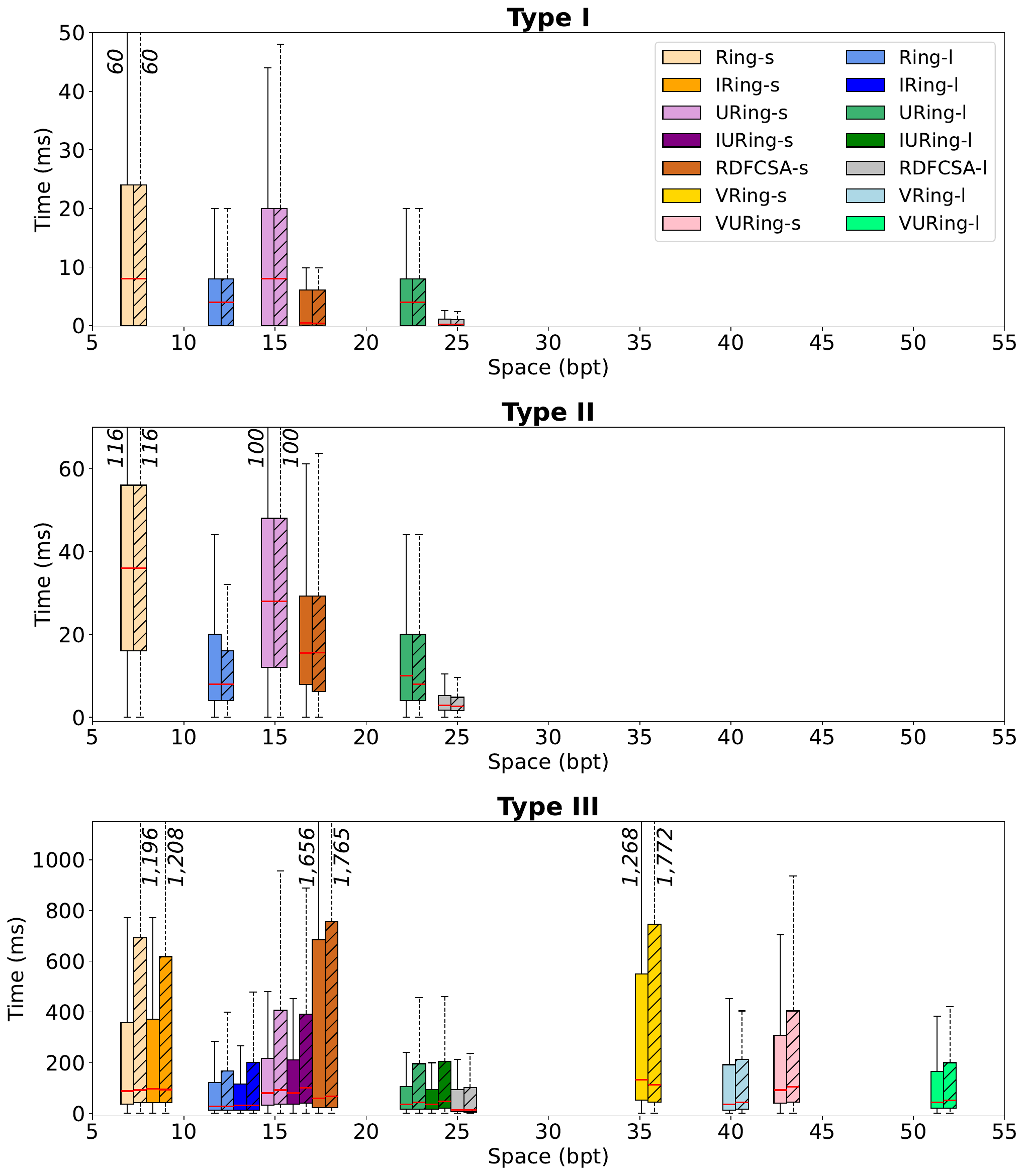}
    \caption{Time distribution, in milliseconds, per query type, limiting outputs to 1000 results. The figures show boxplots for the smaller variants, marking the median inside. Horizontal positions are slightly shifted for visibility; see Table~\ref{tab:limit} for the exact space values. Hatched/empty boxes refer to global/adaptive VEOs. The numbers on top indicate where the capped whiskers reach.}
    \label{fig:limit}
\end{figure}

\subsubsection{The general picture}

It is immediately evident that adaptiveness and the use of the refined estimator is always a good strategy, especially in terms of robustness: average times and timeouts are considerably reduced in all cases. Adaptiveness speeds up all \Ring variants by a factor of 2--11, whereas the refined estimation further speeds up the adaptive variants by a factor of 1.7--2.5.
The dominating strategies, each with its own space-time niche and all using the adaptive variant, are:
\begin{itemize}
    \item The tiny variant: \IRing-small uses just 7.30 bpt (nearly half of a plain storage of the triples), and solves queries with a median of 27 msec and an average of 0.70 sec.
    \item The small variant: \IRing-large uses nearly the space of the triples in plain form (12.15 bpt) and solves queries with a median of 8 msec and an average of 0.18 sec.
    \item The medium variant: \Rdfcsa-large roughly doubles that space (23.54 bpt) and reduces the median to nearly 2 msec. It also offers much better average times on simple queries and with unlimited number of results, as seen later. \IURing-large, using the same space as \Rdfcsa-large, has a worse median but a better average time, 0.16 sec. It is not so interesting, however, in comparison with \IRing-large, which uses half the space and is only marginally slower on average. 
\end{itemize}

\noindent{\em \URing variants.} The one-directional ring, \URing, improves the times of \Ring by 10\%--50\% depending on the variant, in exchange for doubling its space. The largest improvements turn out not to be so relevant, however, because they correspond to the small \URing version, which loses by 20\%--50\% to the large corresponding \Ring version, while using about the same space. The large \URing versions, on the other hand, are only 10\% faster than their corresponding \Ring version, while doubling their space. Using similar space as \URing, \Rdfcsa-large turns out to be more interesting in that it offers more stable times, with a median of 2 instead of 8. \Rdfcsa-small, instead, is not competitive.

 \medskip

\noindent{\em \VRing variants.}
It is also apparent that computing $w_{ij}$ as the number of leaf descendants for choosing VEOs using Eq.~(\ref{eq:wj}) performs much better than the original formula \cite{HRRSiswc19} that uses the number of children of the node: the \VRing variants are much larger and slower than their corresponding \Ring counterparts. 
In the case of adaptive \VRing, where the VEO is recomputed for every binding, this is worsened by the fact that computing $w_{ij}$ on the \VRing takes $O(\log n)$ time, as opposed to $O(1)$ on the \Ring.

\medskip

\noindent{\em Classical systems.}
The best performing classical systems are generally wco: MillenniumDB and (way behind) Jena LTJ, which use about 13 times more space than our ``small'' variant (\IRing-large). MillenniumDB is almost twice as fast as \IRing-large on the average, but has a higher median. This suggests it incurs a base cost of a few tens of milliseconds for every query, even the easy ones. No other classical index (including the non-wco ones) is competitive with our compressed variants.

\subsubsection{Query types and space-time tradeoffs}

Fig.~\ref{fig:limit} shows the time distributions on each query type.
As expected, adaptive query plans and refined estimation of the intersections make no difference with respect to global ones in queries of type I and II, except for a few small gaps. Instead, they make a big difference in queries of type III, more in terms of robustness (lower average, less dispersion) than in the medians. The refined estimation of intersections has, as we have seen, a large impact on averages, but is not noticeable in terms of distribution. This shows that the refinement is mostly useful to avoid few, but very high, query times that are produced when using the coarser measure.

Using twice the space of \IRing-large, \Rdfcsa-large performs better, most clearly on the simpler query types, where the time to extract the output tuples dominates. On those queries \Rdfcsa-large exploits its faster access to the data. Its smaller version, \Rdfcsa-small, instead, is dominated by \IRing-large on queries of type II and III (but not on type I).

The table on top of the figure shows only the best performing variants, which are all adaptive. It clearly shows how times decrease steadily as we use more space. The exception is that, on type-III queries, \IRing-large fares better than \Rdfcsa-large. In this type of queries the query resolution strategy is more important than the mere time to access the data.

\begin{figure}[t]
        \centering                \includegraphics[width=0.8\textwidth]{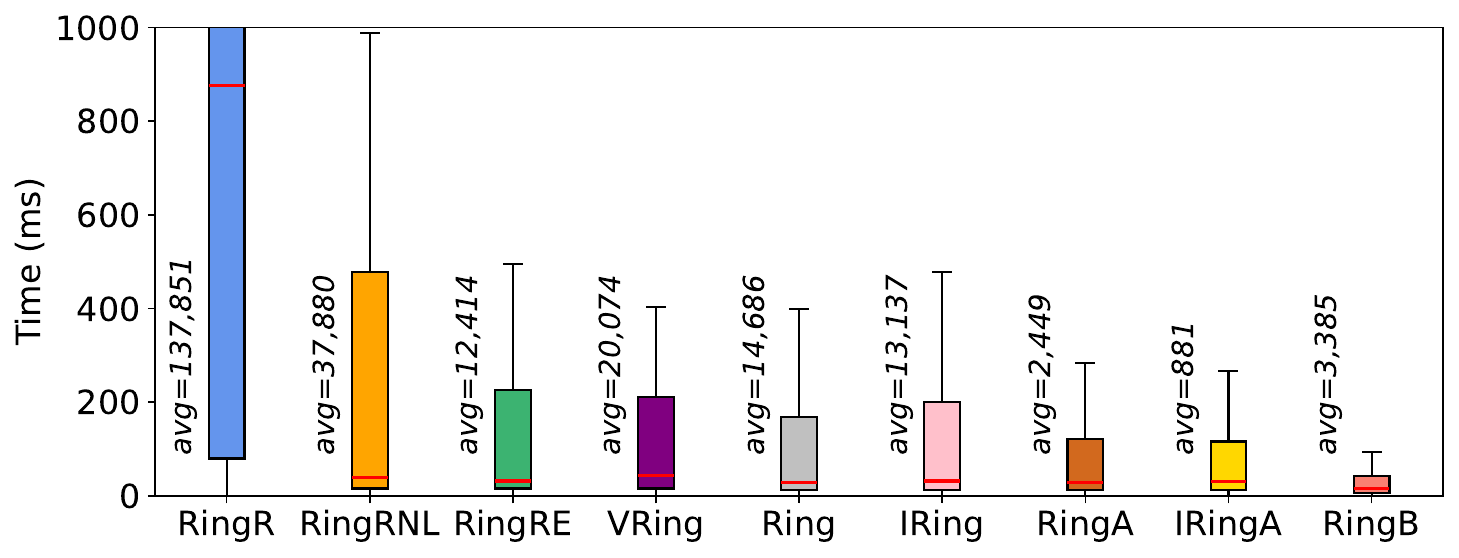}
        \caption{Comparing the (global) VEOs of \VRing, \Ring, and \IRing on queries of type III limited to 1000 results, with variants that choose the VEO at random or optimally.}
        \label{fig:veos}
\end{figure}

\subsubsection{Variable elimination orders}


As already noted, 
the \VRing variants increase the space while worsening query times. This shows up especially on the average times of Table~\ref{tab:limit}, where the \VRing adaptive variants are 2--3 times slower than their corresponding \Ring variants. In part this is due to the $O(\log n)$ time invested in computing the number of children, what can be confirmed by the fact that, on global VEOs where this computation is done only a few times, the \VRing variants are ``only'' 20\%--50\% slower. But still, with global VEOs, where this $O(\log n)$ burden is insignificant, the results show that the \Ring variants generate better VEOs than their \VRing counterparts, apart from computing them faster. The fact that the medians and boxplot distributions are much closer than the averages show that, while most generated VEOs are similar, \Ring avoids very bad VEOs that \VRing sometimes generates.

This leads to the question of how good is the original strategy described in Section~\ref{sec:veo}. To answer it, we created \Ring variants that choose the VEO at random, in nonadaptive form for simplicity, and compare them in Fig.~\ref{fig:veos}. The variant {\Ring}R, which uses a completely random order, is not competitive at all. {\Ring}RNL, which leaves the lonely variables to the end, does much better, showing the convenience of this strategy. The results improve even more on {\Ring}RE, which in addition avoids eliminating variables that are disconnected from previously eliminated ones, if possible. Note that this is just like \Ring, yet using random values to estimate the weights $w_{ij}$. \VRing distributes only slightly better than such a random estimation (and, actually, worsens a lot on the average), but \Ring, which uses the number of leaf descendants to compute $w_{ij}$, performs noticeably better. \IRing, which refines the estimation of the intersections, performs similarly to \Ring when using a global VEO. 
This changes on the adaptive versions: {\Ring}A sharply outperforms \Ring, and {\IRing}A further outperforms {\Ring}A.

An important question here is how much more margin for improvement do we have by choosing VEOs. To partially answer it, we executed \Ring with {\em all the possible global VEOs}, and chose the best time for each query to create an ideal variant called {\Ring}B. As the number of orders to try is the factorial of the number of variables, we reduced the search space by considering only the non-lonely variables and forcing the others to be connected with some eliminated variable in some triple if possible, as \Ring does; we believe the best time should always be within that search space. Further, we did not optimize those queries with 7 or more variables; we just used the time obtained by \Ring on those. The  times of {\Ring}B are then an {\em upper bound} to the best times \Ring could possibly obtain by choosing a good VEO (note that we also leave out adaptive orders). Even so, Fig.~\ref{fig:veos} shows that {\Ring}B sharply outperforms \IRing, our best global VEO, by a factor of about 4 on the average and 2 in the median. {\Ring}B actually outperforms our best adaptive VEO, {\IRing}A, by a factor of about 2 in the median, but interestingly, {\IRing}A is almost 4 times faster on the average. This shows that, in many cases, the adaptive VEOs outperform {\em the best possible} global VEO. On the other hand, the experiment shows that it is still possible to improve a lot upon our current VEOs.
    
\subsubsection{Not limiting the number of results}

\begin{figure}[!t]
\begin{center}
    \includegraphics[width=0.6\textwidth]{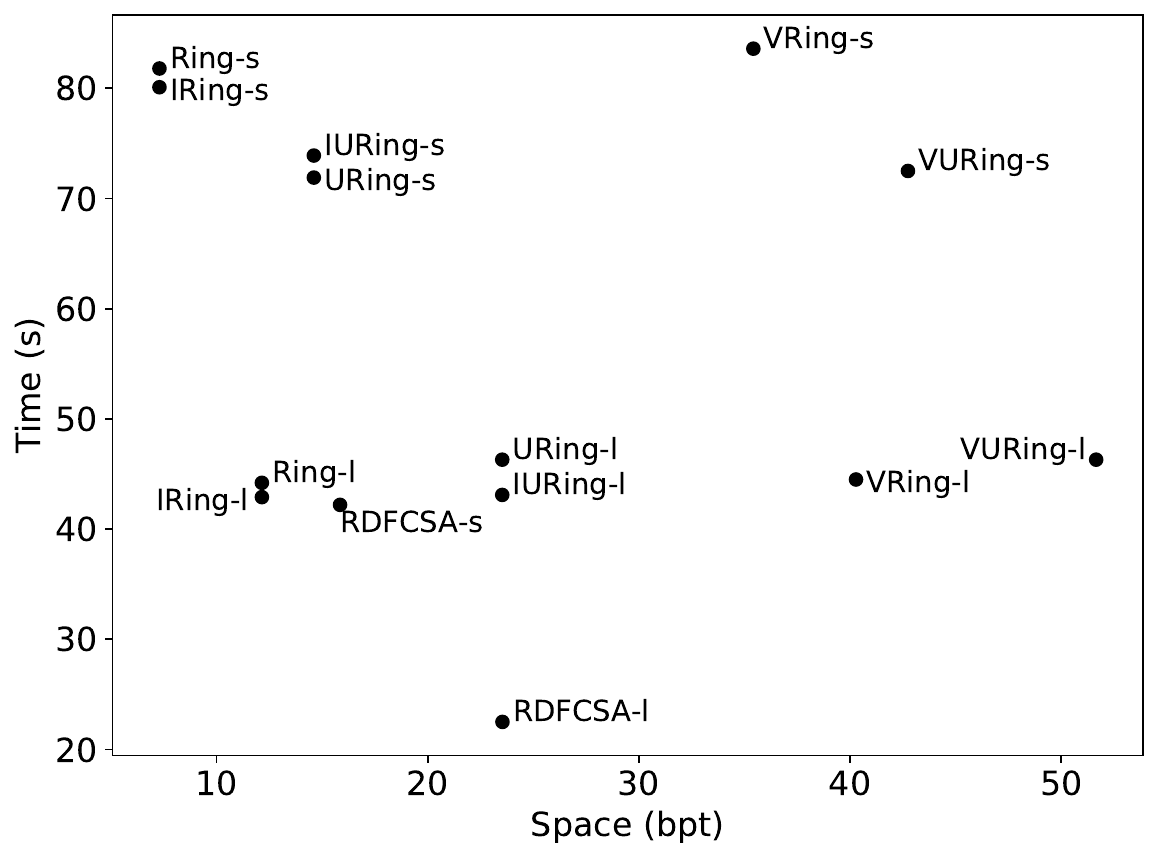}
    \caption{Index space and the averaged query times of the adaptive variants in seconds, not limiting outputs. Suffixes \textit{s} and \textit{l} mean \textit{small} and \textit{large}, respectively.}
    \label{fig:averages-nolimit}
\end{center}
\end{figure}

The case without limits in the number of answers is shown in Table~\ref{tab:nolimit}, where for succinctness we left only the best performing of the classical indices. Fig.~\ref{fig:averages-nolimit} shows the space-time tradeoffs with respect to the average adaptive times.

Although the times are much higher and thus the scale measures seconds, the dominant variants are the same as before. 
An important difference, however, is that adaptiveness and refined estimation of intersections now have little impact on the times. One reason for this is that now the cost to report so many results dominates the overall query time, thereby reducing the relative impact of using better or worse techniques to produce them. Indeed, our times limited to 1000 results suggest that adaptive VEOs produce results {\em sooner} along the query process than global VEOs. 


\begin{table*}
     \centering
        \begin{tabular}{lrrrrrrrrr}
          \toprule
            System      & Space  & \multicolumn{2}{c}{Average} & \multicolumn{2}{c}{Median} & \multicolumn{2}{c}{Timeouts}\\
           & (bpt) & Gl & Ad & Gl & Ad & Gl & Ad \\
            \midrule
            \Ring-small               &  7.30  & 83.6 & 81.8 & 2.9 & 2.9 & 101 & 99 \\
            \IRing-small               &  7.30  & 82.3 & 80.1 & 2.7  & 2.7 & 99 & 94 \\
            \Ring-large               & 12.15  & 46.8 & 44.2 & 0.9 & 0.9 & 59 & 53 \\
            \IRing-large               & 12.15  & 45.4 & 42.9 & 0.8 & 0.8 & 58 & 54 \\
            \URing-small                       & 14.61  & 75.1 & 71.9 & 2.2 & 2.0 & 94 & 86\\
            \IURing-small                       & 14.61  & 72.7 & 73.9 & 2.0 & 2.0 & 89 & 90\\
            \URing-large                       & 23.53  & 47.4 & 46.3 & 1.0 & 1.0 & 59 & 58 \\
            \IURing-large                       & 23.53  & 45.2 & 43.1 & 0.8 & 0.8 & 58 & 55 \\
            \VRing-small                       & 35.42  & 84.6 & 83.6 & 2.9 & 2.9 & 99 & 97\\
            \VRing-large                       & 40.28  & 45.2 & 44.5 & 0.9 & 0.9 & 55 & 53 \\
            \VURing-small                      & 42.74  & 75.1 & 72.5 & 2.2 & 2.1 & 89 & 85\\
            \VURing-large                      & 51.65  & 46.7 & 46.3 & 1.0 & 1.0 & 55 & 56 \\
            \Rdfcsa-small                      & 15.85  & 43.3 & 42.2 & 0.9 & 0.9 & 44 & 43\\
            \Rdfcsa-large             & 23.54  & 22.5 & 21.3 & 0.2 & 0.2 & 31 & 26\\
            \midrule
            MillenniumDB                &  156.78  & \multicolumn{2}{c}{12.0} & \multicolumn{2}{c}{0.05} & \multicolumn{2}{c}{~\,16} \\
            \bottomrule
        \end{tabular}
    \caption{Space and query times (in sec) of all the systems, with Gl(obal) and Ad(aptive) VEOs, not limiting the results. Timeouts count queries exceeding 10 min.}
\label{tab:nolimit}
\end{table*}

\begin{table*}[t]
    \centering
    \begin{tabular}{l@{~~~}r@{~~~}r@{~~~}r@{~~~}r@{~~~}r@{~~~}r@{~~~}r}
    \toprule
      System & Space   & \multicolumn{2}{c}{Type I~~~} & \multicolumn{2}{c}{Type II~~~} & \multicolumn{2}{c}{Type III~~~} \\
             &   (bpt)            & Avg & Med & Avg & Med & Avg & Med \\
      \midrule
      \IRing-small   &   7.30  & 25.5 & 0.100 & 105.4 & 7.80 & 150.4 & 21.89   \\
      \IRing-large   &  12.15 & 10.3 & 0.034 & 52.1 & 2.23 & 103.1 & 5.85   \\
      \IURing-large   &  23.53 & 10.2 & 0.036 & 51.8 & 2.53 & 104.7 & 7.14   \\
      \Rdfcsa-large &  23.54 & 3.1 & 0.004 & 24.5 & 0.74 & 60.6 & 1.84  \\
      \bottomrule
    \end{tabular}
    \caption{The best performing indices, separated by query type, without limiting the results. Times are given in seconds.}
    \label{tab:resumen}
\end{table*}

The fact that the enumeration of results dominates makes \Rdfcsa-large the clear winner of ``medium'' size, not only on the median but also on the average, where it outperforms \IRing-large by a factor of two. MillenniumDB also benefits from reporting many results, because of its locality-friendly data layout. It outperforms \IRing-large by a factor of 3.5 and \Rdfcsa-large by a factor of 1.8 on the average. On the medians, MillenniumDB is 16 times faster than \IRing-large and 4 times faster than \Rdfcsa-large.

Table~\ref{tab:resumen} summarizes the main statistics on the best performing variants, separated by query type. It can be seen that \Rdfcsa-large now clearly outperforms all the \IRing variants even on type-III queries.

\no{
\begin{table*}[t]
\centering
\begin{tabular}{l|r@{~~}c@{~~}c@{~~}c@{~~}c@{~~}c@{~~}c@{~~}c@{~~}c@{~~}c@{~~}c@{~~}c}
System        & Space & $\mathsf{1\textsf{-}tree}$ & $\mathsf{2\textsf{-}tree}$ 
    &  $\mathsf{2\textsf{-}comb}$ 
    & $\mathsf{3\textsf{-}path}$ & $\mathsf{4\textsf{-}path}$  
    & $\mathsf{2\textsf{-}3\textsf{-}lolli}$ & $\mathsf{3\textsf{-}4\textsf{-}lolli}$ 
    & $\mathsf{3\textsf{-}clique}$ & $\mathsf{4\textsf{-}clique}$ 
    & $\mathsf{3\textsf{-}cycle}$ &$\mathsf{4\textsf{-}cycle}$ \\ 
\hline
\Ring-small       &  5.52 & $71.0\, \text{E--}{4}$ & $296\,\text{E--}{4}$ & $54.6\,\text{E--}{4}$
    & $90.7\,\text{E--}{5}$ & $365\,\text{E--}{4}$ 
    & $5.89$ & $535$ 
    & timeout & timeout 
    & timeout & timeout \\
\Ring-large       &  7.59 & $28.6\,\text{E--}{5}$ & $66.4\,\text{E--}{4}$ & $15.2\,\text{E--}{4}$
    & $30.1\,\text{E--}{5}$ & $101\,\text{E--}{4}$ 
    & $1.61$ & $135$
    & timeout & timeout
    & timeout & timeout \\
\Rdfcsa-large     & 14.54 & $8.87\,\text{E--}{5}$ & $24.3\,\text{E--}{4}$ & $5.32\,\text{E--}{4}$
    & $15.1\,\text{E--}{5}$ & $39.5\,\text{E--}{4}$ 
    & $0.405$ & $39.1$ 
    & timeout & timeout 
    & 1636 & timeout \\
\CompactLTJstar   &  6.46 & $4.57\,\text{E--}{5}$ & $\mathbf{6.44\,\textbf{E--}{4}}$ & $\mathbf{1.46\,\textbf{E--}{}4}$
    & $8.89\,\text{E--}{5}$ & $7.14\,\text{E--}{4}$ 
    & $\mathbf{0.116}$ & $\mathbf{10.4}$ 
    & 565 & timeout 
    & 457 & timeout \\
\UnCompactLTJstar &  6.55 & $\mathbf{2.83\,\textbf{E--}{5}}$ & $\mathbf{6.49\,\textbf{E--}{4}}$ & $\mathbf{1.50\,\textbf{E--}{4}}$
    & $\mathbf{5.11\,\textbf{E--}{5}}$ & $\mathbf{6.76\,\textbf{E--}{4}}$ 
    & $\mathbf{0.113}$ & $\mathbf{10.2}$ 
    & 558 & timeout 
    & 452 & timeout \\
\hline
Graphflow & 13.54 & & & & & & & & $83.3$ & $975$ & $80.9$ & timeout \\
ADOPT-1           & 20.09 & $0.817$ & $1.67$ 
    & $1.03$ & $1.28$ & $1.15$
    & $6.52$ & timeout 
    & 1337 & timeout
    & 885 & timeout\\
ADOPT-70          & 20.09 & $0.837$ & $1.75$ & $1.21$
    & $1.12$ & $1.56$ 
    & $3.60$ & $105$ 
    & 105 & timeout 
    & 106 & timeout \\
EmptyHeaded       & 28.65 & $5.76\,\text{E--}{5}$ & $1.32$ & $9.68\,\text{E--}{4}$
    & $45.5\,\text{E--}{5}$ & $0.506$ 
    & $11.0$ & $315$ 
    & $\mathbf{14.0}$ & $\mathbf{326}$ 
    & $\mathbf{13.1}$ & $\mathbf{1006}$ \\
\end{tabular}
\caption{Space in bpt and median time in seconds (timeout is 1800) for various systems on graph $\mathtt{soc\texttt{-}LiveJournal1}$.}
\label{tab:beyond-wco}
\end{table*}

\subsection{Beyond wco systems} \label{sec:beyond-wco}

The alternative systems we have compared either are not wco, or use the basic LTJ with some global VEO. In this section we compare our new compact indices, with the new VEOs we have developed on them, against systems that use more sophisticated query resolution strategies:
\begin{itemize}
\item Graphflow \cite{graphflow}: A graph query engine that indexes property graphs using in-memory sorted adjacency lists and supports hybrid plans blending wco and pairwise joins. 
\item ADOPT \cite{adopt}: The first wco algorithm using adaptive VEOs on LTJ. It uses exploratory search and reinforcement learning to find near-optimal orders, using actual execution times as feedback on the suitability of orders. We include variants using one and 70 threads.
\item EmptyHeaded \cite{EmptyHeaded}: An implementation of a more general algorithm than LTJ, which applies a generalized hypertree decomposition~\cite{gottlob1999hypertree} on the queries and uses a combination of wco algorithms \cite{NPRR12} and Yannakakis' algorithm~\cite{yannakakis}. Triples are stored in 6 tries (all orders) in main memory.
\end{itemize}

Those systems use too much memory on our Wikidata graph. For example, Graphflow stores one structure per predicate, which makes it usable with few predicates only: on a subset containing $< 10\%$ of our Wikidata graph \cite{AHNRRSsigmod21}, it failed to build even in a machine with 730 GB of Java heap space. ADOPT did not build correctly either. EmptyHeaded runs but it uses 1810 bpt, over 10 times more than Jena LTJ.

In this section we compare them over an even smaller graph used in previous work \cite{NABKNRR15}, $\mathtt{soc\texttt{-}LiveJournal1}$, the largest from the Stanford Large Network Dataset Collection~\cite{stanford}, with $68{,}993{,}773$ {\em unlabeled} edges. 
We test different query shapes (see previous work for a detailed description \cite{NABKNRR15}) including trees ($\mathsf{1\textsf{-}tree}$, $\mathsf{2\textsf{-}tree}$, $\mathsf{2\textsf{-}comb}$), paths ($\mathsf{3\textsf{-}path}$, $\mathsf{4\textsf{-}path}$), paths connecting cliques ($\mathsf{2\textsf{-}3\textsf{-}lollipop}$, $\mathsf{3\textsf{-}4\textsf{-}lollipop}$), cliques ($\mathsf{3\textsf{-}cliques}$, $\mathsf{4\textsf{-}cliques}$), and cycles ($\mathsf{3\textsf{-}cycles}$, $\mathsf{4\textsf{-}cycles}$). We include 10 queries for each tree, path, and lollipop, and 1 for each clique and cycle. This is the same benchmark used for ADOPT \cite{adopt}, except that we do not force the clique and cycle variables to be different, and we choose for the constant any random value such that the query has occurrences.
%
%
%
%
%
We set a 30-minute timeout and do not limit the number of results. 

Since there are no labels, the \Ring and \Rdfcsa variants need not store the columns nor the part of $\Psi$ for predicates, and the others store only two orders, $\textsc{pso}$ and $\textsc{pos}$. Graphflow is tested on the cliques and cycles only because the implementation does not support constants in the BGPs.

Table~\ref{tab:beyond-wco} shows spaces and times. Interestingly, \CompactLTJstar and \UnCompactLTJstar get close to the least space, being only larger than \Ring-small. Graphflow, ADOPT and EmptyHeaded use 2, 3, and over 4 times more space, respectively. The tree and path queries are solved in microseconds by \CompactLTJstar/\UnCompactLTJstar, while the slowest \Ring/\Rdfcsa is up to 25--50 times slower. ADOPT is 4--5 orders of magnitude slower in these queries (parallelization does not help in this case). EmptyHeaded is from twice as slow to 3--4 orders of magnitude slower.

The lollipop shapes are harder, but \CompactLTJstar/\UnCompactLTJstar still handle them in at most 10 seconds, being 1--2 orders of magnitude faster than ADOPT and EmptyHeaded. The parallel ADOPT is 3 times faster than EmptyHeaded in these shapes. 

EmptyHeaded finally takes over on the hardest shapes, cliques and cycles, where it is 3--6 times faster than Graphflow, 7--8 times faster than the parallel ADOPT, and 35--40 times faster than \CompactLTJstar/\UnCompactLTJstar. We note that the latter are still twice as fast as sequential ADOPT.
}

\no{
\subsection{\textcolor{red}{Graph patterns benchmark}} \label{sec:patterns}

We further study the performance of the dominant compact and classical indices, now classified by the topology of the queries. We also include \QDag, which uses 10.41 bytes per triple, standing between \Ring-small and \Ring-large. Fig.~\ref{fig:exp2} shows how the query times distribute across the 17 topologies, limiting results to 1000. 

The relative performances of \Ring-small, \Ring-large, \Rdfcsa-large, \CompactLTJ, and \UnCompactLTJ stay similar across all query shapes, as expected from them all using essentially the same \LTJ algorithm on different representations. They all fare better on the acyclic than on the cyclic queries (e.g., the \Ring-large medians are within 8--16 msec on acyclic versus 22--46 msec on cyclic queries). 

\QDag, instead, performs differently, because it uses another algorithm. In line with previous work \cite{ANRRtods22}, \QDag excels on the small cyclic queries \textsf{Tr1}, \textsf{Tr2}, and \textsf{S1}, where it approaches the performance of the (twice as large) \Rdfcsa-large. It also does relatively well in the smallest acyclic queries, \textsf{P2}, \textsf{T2}, and \textsf{Ti2}, where it nearly reaches the performance of \Ring-small or \Ring-large. This is explained because the time complexity of \QDag is multiplied by $2^d$, where $d$ is in our case the number of nodes in the graphs of Fig.~\ref{fig:wikidata-subgraph-query-patterns} \cite{ANRRtods22}.

MillenniumDB, on the other hand, stands out for its stability. Its times are around 20 msec, with low variance independently of the query shape. While its performance is worse than that of \Ring-large on the acyclic queries, MillenniumDB outperforms \Ring-large (but not \Rdfcsa-large) on the cyclic queries. The reason may lie in the (global) VEO used by MillenniumDB, which is similar to the one of \Ring, but MillenniumDB stores additional data that enables a finer-grained estimation of the expected number of children as variables are eliminated \cite{VR+23}: it stores the number of different subjects and objects for each distinct predicate. This turns out to be particularly useful for the regularity of the WGPB benchmark, where all predicates are constant and all subjects and objects are variable. Still, it would be interesting to study the combination of its more sophisticated algorithm for global VEOs with the compact representations, on real query logs. 
 
\begin{figure}
    \centering
    \includegraphics[width=0.4\textwidth]{figures/exp2-box-1.pdf}
    \includegraphics[width=0.4\textwidth]{figures/exp2-box-2.pdf}
\includegraphics[width=0.4\textwidth]{figures/exp2-box-3.pdf}
\includegraphics[width=0.4\textwidth]{figures/exp2-box-4.pdf}
\includegraphics[width=0.4\textwidth]{figures/exp2-box-5.pdf}
\includegraphics[width=0.4\textwidth]{figures/exp2-box-6.pdf}
    \caption{Time distribution, in seconds, classified by the query patterns of Fig.~\ref{fig:wikidata-subgraph-query-patterns}. \textcolor{red}{Renombrar a CLTJ* y UnCLTJstar}}
    \label{fig:exp2}
\end{figure}
}

\no{
\begin{table}[]
\caption{ Data + Index space in bytes per triple and query times (secs), on the Wikidata sub-
graph.}
\label{table:bytes-per-triple-wikidata-subgraph}
\begin{tabular}{lll}
\hline
System (Data + Indices) & Space  & Avg \\
\hline
QDags                   & 4.9      & 16.04\\
Ring                    & 11.16    & 0.018\\
Ring Adaptive           & 11.16    & 0.020\\
RDFCSA                  & 15.04    & 0.012\\
URing                   & 22.32    & 0.020\\
URing Adaptive          & 22.32    & 0.020\\
Ring Muthu              & 26.24    & 0.018\\
Ring Adaptive Muthu     & 26.24    & 0.013\\
CompactLTJ              & 43.74    & 0.008 \\
URing Muthu             & 53.36    & 0.016\\
URing Adaptive Muthu    & 53.36    & 0.012\\
\hline
MillenniumDB            & 152      & 0.012\\
Virtuoso                & 104.89   & 1.135 \\
Blazegraph              & 99.86    & 1.709 \\
RDF-3X                  & 107.67   & 0.182 \\
Jena LTJ                & 144.64   & 0.059 \\
Empty Headed            & 1,809.84  & 0.118\\
\hline
\end{tabular}
\end{table}

\begin{figure*}[t]
\includegraphics[width=1.2\textwidth]{figures/tradeoff_bytes_per_triple_median.pdf}
\caption{Tradeoff Median / bytes-per-triple }
\label{fig:wikidata-subgraph-tradeoff}
\end{figure*}

\begin{figure*}[t]
\includegraphics[width=.9\textwidth]{figures/boxplots_by_type_ring_variants_only.pdf}
\caption{Different Ring variants query times comparison (in seconds).}
\label{fig:boxplot_ring_benchmark}
\end{figure*}

\begin{figure*}[t]
\includegraphics[width=.9\textwidth]{figures/boxplots_by_type_all_indexes.pdf}
\caption{Comparison of query times per index (in seconds).}
\label{fig:boxplot_all_indexes_benchmark}
\end{figure*}

\blockcomment{
\begin{figure*}[t]
\captionsetup[subfloat]{labelformat=empty}
\subfloat[]
  {
    \includegraphics[width=.9\textwidth]{figures/boxplots_by_type_all_row_1.pdf}
  }
  \\
\subfloat[]
  {
    \includegraphics[width=.9\textwidth]{figures/boxplots_by_type_all_row_2.pdf}
  }
  \\
\subfloat[]
  {
    \includegraphics[width=.9\textwidth]{figures/boxplots_by_type_all_row_3.pdf}
  }
\caption{Comparison of query times per index (in seconds).}
\label{fig:boxplot_cds_db_engines_benchmark}
\end{figure*}
}
}


\section{Conclusions}

We have introduced new compact indices, combined with novel query resolution strategies, to solve Basic Graph Patterns on graph databases using Leapfrog TrieJoin (\LTJ), the leading worst-case-optimal (wco) multijoin algorithm. Concretely:

\begin{itemize}
    \item We uncover a \emph{space-time tradeoff} formed by compact indices. The Pareto-optimal variants use from about 0.6 to 2.0 times the space needed to store the triples in raw form. These compressed indices are outperformed only by the classic LTJ implementation in MillenniumDB \cite{VR+23}, which uses 14 times the space needed to store the triples. 
    \item We combine those new indices with \emph{adaptive variable elimination orders}, in contrast with the global orders in use. We show that adaptively choosing the next variable to eliminate along the query process yields results much sooner: our adaptive query plans are up to 5 times faster to obtain the first 1000 results, and they even outperform in many cases the {\em best possible} non-adaptive plan. The time to obtain all the results, instead, is nearly the same. 
    \item We show that using the total number of leaves descending from an \LTJ trie node yields much better variable elimination orders compared with the classic measure of the number of children of the node: while the latter approximates the cost of performing the next intersection, the former better estimates the whole future cost. Our better estimation speeds up query resolution by a factor over 2, and almost 8 when combined with adaptive plans.
    \item We also show that the estimation of the intersection size can be refined by using features that are unique of our compact representation, which further speeds up the adaptive query times by a factor of up to 2.4. Further, we show that there is much space for improvement in terms of choosing a good variable ordering when generating query plans.
\end{itemize}

Overall, our new representations outperform the original \ring  by a factor up to 13 to produce the first 1000 results, while using the same space, and by 2 overall using about twice the space; this doubled space is still several times less than those of classical indices. Classical wco and non-wco indices are (often sharply) outperformed by our fastest variants. Only one of those, using 4 times the space of our largest relevant variant, outperforms it by a factor 
2 on the average. 
    
We remark that our compact indices run in main memory and would not be disk-friendly. While their compactness make them fit in memory for larger datasets, a relevant future work direction is to design compact representation formats for disk or distributed memory, where compactness translates into fewer I/Os or communication at query resolution time.

Another limitation of our compact indices is that they do not yet support updates. A way to support updates is to replace the wavelet tree bitvectors of all the \ring structures by dynamic variants \cite{MN08}. In this case, a node or triple can be inserted or deleted in the graph in $O(\log n \log\sigma)$ time. This, however, multiplies all the operation times by $O(\log n)$, which is a high price especially if we consider that updates are not only infrequent in many use cases, but also that queries require many more accesses than updates. A recent development of dynamic bitvectors that are sensitive to the frequency of queries versus updates \cite{Nav24} yields a promising implementation of dynamism in our scenario: we anticipate almost no increase in query times by supporting efficient updates in the graph.

\bmhead{Data availability statement} 
The data and source code that support the findings of this study are openly available in zenodo
at \url{https://zenodo.org/records/13141588}  and \url{https://zenodo.org/records/13142779}, respectively.

\bmhead{Acknowledgements}

{
 Supported by ANID -- Millennium Science Initiative Program -- Code ICN17\_002, Chile. A.F. and A.G. are funded in part
 by MCIN/AEI/10.13039/501100011033:  grant PID2020-114635RB-I00 (EXTRACompact); 
 by MCIN/AEI/10.13039/ 501100011033 and EU/ERDF "A way of making Europe": PID2021-122554OB-C33 (OASSIS) and PID2022-141027NB-C21 (EARTHDL);
 by MCIN/AEI/10.13039/501100011033 and “NextGenerationEU”/ PRTR: grants TED2021-129245B-C21 (PLAGEMIS), PDC2021-120917-C21 (SIGTRANS)
 and by GAIN/Xunta de Galicia: GRC: grants ED431C 2021/53, and CIGUS 2023-2026. G.N. is funded in part by Fondecyt Grant 1-230755, Chile.
}



\bibliography{sn-bibliography}
\balance


\end{document}